\documentclass[letterpaper,11pt]{article}
\usepackage{amssymb}
\usepackage{amsmath}
\usepackage{amstext}
\usepackage{graphicx,epsfig}
\usepackage{epsfig}
\usepackage{verbatim} 
\usepackage{fancybox}
\usepackage{color}
\usepackage{ulem}
\usepackage{enumitem}
\usepackage{subfigure}
\usepackage{bbm}
\usepackage{parskip}
\usepackage{dsfont}
\usepackage[numbers,sort&compress]{natbib}

\newcommand{\be}{\begin{equation}}
\newcommand{\ee}{\end{equation}}
\newcommand{\bea}{\begin{eqnarray}}
\newcommand{\eea}{\end{eqnarray}}
\newcommand{\beas}{\begin{eqnarray*}}
\newcommand{\eeas}{\end{eqnarray*}}

\def\({\left(}
\def\){\right)}

\newcommand\tg{{\tilde g}}

\newcommand\bQ{{\bar Q}}

\def\gsim{ \lower .75ex \hbox{$\sim$} \llap{\raise .27ex \hbox{$>$}} }
\def\lsim{ \lower .75ex \hbox{$\sim$} \llap{\raise .27ex \hbox{$<$}} }

\newcommand\pd{\partial}
\newcommand\mpl{M_{\rm Pl}}
\newcommand\rmd{{\rm d}}

\newcommand\na{\nabla}
\newcommand\la{\langle}
\newcommand\ra{\rangle}

\newcommand\nn{\nonumber}

\usepackage{xcolor,colortbl}

\begin{document}
\def\thefootnote{\fnsymbol{footnote}}

\begin{center}
\Large{\textbf{A Universe Without Dark Energy: Cosmic Acceleration from Dark Matter-Baryon Interactions}} \\[0.5cm]
 
\large{Lasha Berezhiani$^{1}$, Justin Khoury$^{2}$ and Junpu Wang$^{3,4}$}
\\[0.5cm]

\small{
\textit{$^{1}$ Department of Physics, Princeton University, Princeton, NJ 08540\\ $^{2}$ Center for Particle Cosmology, University of Pennsylvania, Philadelphia, PA 19104\\
$^{3}$ Department of Physics and Astronomy, Johns Hopkins University, Baltimore, MD 21218\\
$^{4}$ Department of Physics, Yale University, New Haven, CT 06511}
}

\vspace{.2cm}

\end{center}

\vspace{.2cm}

\hrule \vspace{0.2cm}
\centerline{\small{\bf Abstract}}
{\small Cosmic acceleration is widely believed to require either a source of negative pressure ({\it i.e.}, dark energy), or a modification of gravity, which necessarily implies new degrees of freedom beyond those of Einstein gravity. In this paper we present a third possibility, using only dark matter and ordinary matter. The mechanism relies on the coupling between dark matter and ordinary matter through an effective metric. Dark matter couples to an Einstein-frame metric, and experiences a matter-dominated, decelerating cosmology up to the present time. Ordinary matter couples to an effective metric that depends also on the DM density, in such a way that it experiences late-time acceleration. Linear density perturbations are stable and propagate with arbitrarily small sound speed, at least in the case of `pressure' coupling. Assuming a simple parametrization of the effective metric, we show that our model can successfully match a set of basic cosmological observables,  
including luminosity distance, BAO measurements, angular-diameter distance to last scattering {\it etc.}  For the growth history of density perturbations, we find an intriguing connection between the growth factor and the Hubble constant. To get a growth history similar to the $\Lambda$CDM prediction, our model predicts a higher $H_0$, closer to the value preferred by direct estimates. On the flip side, we tend to overpredict the growth of structures whenever $H_0$ is comparable to the Planck preferred value. The model also tends to predict larger redshift-space distortions at low redshift than $\Lambda$CDM.}
\vspace{0.3cm}
\noindent
\hrule
\def\thefootnote{\arabic{footnote}}
\setcounter{footnote}{0}

\section{Introduction}

There is a folk theorem which, roughly speaking, states that late-time cosmic acceleration can arise in only one of two ways: either it is due to dark energy, {\it i.e.}, a source of negative pressure, such as a cosmological constant or `quintessence'~\cite{Ratra:1987rm,Wetterich:1987fm,Caldwell:1997ii,Copeland:2006wr}; or it is due to a modification of Einstein gravity~\cite{Joyce:2014kja}, such as in massive gravity,
which necessarily implies new degrees of freedom beyond the standard helicity-2 gravitons. Naturally people have considered hybrid models that do both, {\it e.g.}, dark energy scalar fields interacting with dark matter (DM)~\cite{Amendola:1999er} as well as normal matter, as in chameleon~\cite{Khoury:2003aq,Khoury:2003rn}, symmetron~\cite{Hinterbichler:2010es,Hinterbichler:2011ca}. But it appears that either dark energy or new degrees of freedom is necessary to explain cosmic acceleration. 

In this paper we present a loophole in the theorem. There is a third possibility: cosmic acceleration arising from suitable interactions between DM and baryons, without sources of negative pressure (in the Einstein frame) or new degrees of freedom beyond DM and ordinary matter. The mechanism relies on dark matter and baryons coupling to different
metrics. Our approach is purposely agnostic about the microphysical nature of DM and applies equally well to WIMPs~\cite{Bertone:2004pz}, axions~\cite{Preskill:1982cy,Abbott:1982af}, ultra-light scalar field DM~\cite{Sin:1992bg,Ji:1994xh,Hu:2000ke,Goodman:2000tg,Peebles:2000yy,Silverman:2002qx,Arbey:2003sj,Boehmer:2007um,Lee:2008ux,Lee:2008jp,Dwornik:2013fra,Guzman:2013rua,Hui:2016ltb} or superfluid DM~\cite{Berezhiani:2015pia,Berezhiani:2015bqa,Khoury:2016ehj,Khoury:2016egg}. In particular the coupling described below through an effective metric is above and beyond other allowed WIMP-like or axion-like couplings the DM may have with ordinary matter.  For concreteness we therefore ignore such additional model-independent couplings, since they have negligible impact on late-time cosmology.

It must be emphasized that our model is an effective field theory with high-dimensional operators and as such it does propagate additional degrees of freedom at the cut-off. However, as we will see, the spice of our model lies in the fact that these degrees of freedom are much heavier than the Hubble scale. That is, the late-time cosmology is governed by the dynamics at energy scales much below the cut-off, where a direct coupling between DM and baryons is induced by integrating out the heavy degrees of freedom.

For the moment let us put aside the particle physics model construction at high energy, and focus on the effective field theory for cosmology:
Dark matter couples to an Einstein-frame metric $g_{\mu\nu}$ and experiences a decelerating, approximately matter-dominated expansion up to the present time. Baryons instead couple to a physical (or `Jordan-frame') metric $\tilde{g}_{\mu\nu}$, constructed from $g_{\mu\nu}$ and the physical parameters of the DM component. As discussed in Sec.~\ref{setupsection}, treating DM in the hydrodynamical limit as a perfect and vorticity-free fluid, DM can be described effectively as a $P(X)$ scalar theory, where $X = -g^{\mu\nu}\pd_\mu\Theta \pd_\nu \Theta$. The variables at our disposal for $\tilde{g}_{\mu\nu}$ are the DM 4-velocity $u_\mu = \frac{\partial_\mu \Theta}{\sqrt{-X}}$ and $X$. Its most general form is therefore 

\be
\tilde{g}_{\mu\nu}=R^2(X)\left( g_{\mu\nu}+u_\mu u_\nu \right)-Q^2(X)u_\mu u_\nu\,.
\label{tildegintro}
\ee
Hereinafter we will use letters with (without) tilde to denote quantities in the Jordan (Einstein) frame.
In a microscopic model of DM, one can think of $Q$ and $R$ to be functions of some scalar composite operator made of DM fields, e.g. the energy density. In the present case, these scalar functions $R$ and $Q$ are chosen such that: $i)$ they tend to unity at high DM density, in order to reproduce standard evolution at early times; $ii)$ they grow at late times (roughly, at redshift $\sim$ a few) to generate apparent cosmic acceleration for ordinary matter. Thus at the level of the background evolution it seems straightforward to obtain cosmic acceleration for judicious choice of $R$ and $Q$.
One can even fine-tune these functions to {\it exactly} match the $\Lambda$CDM expansion history, though in our analysis we will consider more general functional forms. See Sec.~\ref{overviewsection} for an overview of the mechanism and Sec.~\ref{Backgroundandperturb} for a discussion of the background evolution. 

It is worth pointing out that incidentally condition $i)$ enforces a screening mechanism, so that the direct coupling between DM and ordinary baryon particles will not result in any violation of the Equivalence Principle. That is, no ``fifth force'' on ordinary matter due to the mediation of DM will be detected in local gravity experiments.
 Our condition $i)$ and $ii)$ on $R$ and $Q$ implies that the direct interaction between DM and baryon particles is not turned on until the ambient DM density is sufficiently low.\footnote{We thank David E.~Kaplan for discussions on this point.} As we will see in Section \ref{integratein} this seemingly counter-intuitive fact arises naturally from a microscopic model.

We would like to stress that our model for late-time cosmology does {\it not} fit into the paradigm of quintessence theories, although there are similar features such as the presence of two metrics. The key difference is that there are no additional light fields in our model. Said differently, cosmic acceleration in our model is due to composite operators made of DM. As we mentioned above, the origin of these composite operators can be traced back to integrating out heavy degrees of freedom.

What about the growth of density perturbations? Because the Einstein-frame scale factor evolves as approximately dust-dominated, $a \sim t^{2/3}$, and since DM couples minimally to this metric, density perturbations grow as in standard CDM $\delta \sim a$. Naively this would seem to rule out the model, since an important consequence of cosmic acceleration is that it slows down the growth of structures, consistent with observations. However in our model the {\it observed} growth rate should be measured relative to the {\it physical} scale factor $\tilde{a} = R a$, resulting in an effective growth function $\sim \frac{a}{\tilde{a}} = R^{-1}$. Thus the same function $R$ that grows at late times to mimic cosmic acceleration also serves to suppress the growth of structures. More physically, this can be understood as a time dilation effect. Although perturbations grow unimpeded in  the Einstein frame, the Einstein-frame universe is younger than the `Jordan-frame' universe experienced by ordinary matter. Hence, from the perspective of ordinary matter, cosmic structures appear less developed than in a pure CDM universe. Thus the {\it observed} growth history matches that of a universe with dark energy, though in general it is not identical to $\Lambda$CDM. 

A critical test for the viability of the model is the stability of linear perturbations. In Sec.~\ref{linpert} we carefully study perturbations, in the limit that modes are well-inside the horizon, 
such that mixing with gravity can be ignored. Even in this simplified regime, because DM and baryons are coupled through $\tilde{g}_{\mu\nu}$, their perturbations are
kinetically-mixed. Perturbative stability requires that kinetic and gradient matrices both be positive-definite, and we find that the resulting conditions on $Q$ and $R$ are easy
to satisfy. A more stringent constraint, however, comes from imposing that the sound speed be sufficiently small, $c_s\ll 1$, to avoid unwanted oscillations in the matter power spectrum~\cite{Sandvik:2002jz,Wang:2013qy}. Of the two propagating scalar modes, we find that one mode propagates with a sound speed that vanishes identically as a result of baryons being pressureless. The sound speed for the second mode is more complicated and depends explicitly on the form of DM-baryon interactions.

For the conformal coupling ($Q = R$), in which case~\eqref{tildegintro} reduces to $\tilde{g}_{\mu\nu} = Q^2 g_{\mu\nu}$, demanding that $Q$ varies sufficiently fast to drive cosmic acceleration generically leads $c_s$ becoming relativistic at late times. This case is therefore phenomenologically disfavored. Thus we are led to consider the `maximally-disformal' or `pressure' coupling $Q = 1$, for which~\eqref{tildegintro} implies that the background metric in the Jordan frame takes the form of ${\rm d}\tilde{s}^2 = -{\rm d}t^2 + R^2a^2(t){\rm d}\vec{x}^2$. In this case we show that the sound speed 
can be made arbitrarily small, as desired. This is the coupling ``de choix" for the rest of our analysis.

Although our scenario does not require any additional degree of freedom beyond DM and ordinary matter, for some applications it is conceptually helpful to ``integrate in"  additional fields. Section~\ref{integratein} provides such a formulation, by introducing a scalar $\phi$ and vector $A_\mu$. In the limit that these fields are very heavy, and therefore approximately auxiliary, their expectation value is fixed by the DM density and 4-velocity respectively as $\phi \sim \rho_{\rm DM}$ and $A^\mu \sim u^\mu$. This formulation is particularly helpful to discuss constraints from direct detection experiments. Treating DM as fermions for concreteness, we find that the DM-baryon coupling reduces to an effective, density-dependent 4-fermion vertex. The effective Fermi constant can be made arbitrarily small as $c_s\rightarrow 0$, and in fact vanishes in the maximally-disformal case $Q = 1$. 

In Sec.~\ref{cosmoobs} we derive the observational predictions for our model and compare the result to the $\Lambda$CDM model. We focus
on the phemenologically-viable maximally-disformal case $Q = 1$, leaving us with a single function $R(\tilde{a})$ to fully specify the model. 
For concreteness in Sec.~\ref{fidmod} we choose a simple, Taylor-series parametrization of this function, in terms of two constants $\alpha$ and $\beta$. We begin by
imposing two conservative restrictions on the $(\alpha,\beta)$ parameter space. First, we demand that the cosmological proper distance
$\tilde{H}_0 d_{\rm P}(\tilde{z})$, where $\tilde{H}_0$ is the observed Hubble constant, agree with the $\Lambda$CDM prediction to within 3\% over the
redshift range $0\leq \tilde{z}\leq 3$. Second, by matching to the angular diameter distance to the cosmic microwave background (CMB), we demand that
our predicted Hubble constant $\tilde{H}_0$ lies between 65 and 75~${\rm km}\;{\rm s}^{-1}{\rm Mpc}^{-1}$. This range is chosen to include, at the lower end,
the Planck best-fit $\Lambda$CDM value~\cite{Aghanim:2016yuo} $H_0^{\Lambda{\rm CDM}} = 66.93\pm 0.62~{\rm km}\;{\rm s}^{-1}{\rm Mpc}^{-1}$,
and, at the upper end, the direct Hubble Space Telescope (HST)~\cite{Riess:2016jrr} measurement of $H_0^{\rm direct} = 73.24 \;\pm\; 1.74~{\rm km}\;{\rm s}^{-1}{\rm Mpc}^{-1}$. 

With these two priors, we then go on to calculate various cosmological observables, including the luminosity distance relation (Sec.~\ref{lumdis}), Baryon Acoustic Oscillations (BAO) (Sec.~\ref{BAO}), and the growth function of density perturbations (Sec.~\ref{growthhistorysection}). In the process we discover an intriguing connection between the growth factor and the Hubble constant. In the region of $(\alpha,\beta)$ parameter space where the predicted $\sigma_8$ is comparable to the Planck best-fit $\Lambda$CDM value of $\sigma_8 = 0.83$, the predicted Hubble constant tends to be on the high side, closer to the direct HST estimate. (Although there is agreement on $\sigma_8$, the quantity $f\sigma_8$ probed by redshift-space distortions, where $f$ is the growth rate, is systematically higher than in $\Lambda$CDM at low redshift.) On the other hand, in the region of $(\alpha,\beta)$ parameter space where $\tilde{H}_0$ is comparable to the Planck preferred value, then we predict higher values of $\sigma_8$ (and $f\sigma_8$), which tends to exacerbate the existing mild tension with weak lensing and cluster counts~\cite{Ade:2015xua}. Thus to get a sensible growth history our model predicts a higher Hubble constant than $\Lambda$CDM, in better agreement with direct estimates. It remains to be seen whether this conclusion holds generally or is specific to the simple $(\alpha,\beta)$ parametrization adopted here.

One observable we do not consider here is the CMB angular power spectrum, as this requires modifying the CAMB numerical code.
The full derivation of the CMB spectrum will be presented elsewhere~\cite{AdamVin}. However, since by design the two metrics
$g_{\mu\nu}$ and $\tilde{g}_{\mu\nu}$ coincide at early times, resulting in decoupled DM-baryon sectors at recombination, we expect negligible
impact on the CMB spectrum on small angular scales. As with dark energy, the main impact on the CMB is felt on large angular scales, through the
integrated Sachs-Wolfe (ISW) effect. In Sec.~\ref{ISWsection} we give a preliminary estimate of the ISW contribution and find that it may
be problematic for our model. Specifically, the predicted ISW signal is strongly scale-dependent and peaks on small scales, which naively
implies a large ISW signal. On the other hand, this may be a good thing --- the observed cross-correlation is larger than the $\Lambda$CDM prediction by about $2\sigma$, {\it e.g.},~\cite{Liu:2015xfa}. This warrants further study.

Our model is not the first attempt to ``unify" DM and DE. The most famous example is the Chaplygin gas~\cite{Bilic:2001cg,Bento:2002ps}, which proposes that DM is a substance with an unusual equation of state $P \sim - \rho^{-\alpha}$. This component therefore behaves as dust ($P\simeq 0$) at high density and as dark energy ($P < 0$) at low density. However, in this model the sound speed $c_s \sim \alpha$ can become significant at late times, resulting in either large oscillations or exponential blow-up in the matter spectrum~\cite{Sandvik:2002jz,Wang:2013qy}. In our case, the DM has $c_s\simeq 0$ at all times (just like CDM), and the matter power spectrum is consistent with observations. Closer in spirit to our model is the `abnormally weighting energy' model~\cite{Alimi:2008ee}, in which DM and baryons couple differently to a Brans-Dicke scalar field. DM sources the time-evolution of the scalar field, which in turn results in the baryon metric undergoing cosmic acceleration. A key difference is that our model does not need any new degrees of freedom, scalar or otherwise, beyond DM and ordinary matter.\footnote{The idea of having different cosmologies in Einstein and Jordan frames has been explored in the context of the early universe cosmology in \cite{Ijjas:2015zma}.}

\section{Set Up}
\label{setupsection}

Our mechanism is most intuitive in the `Einstein frame', where the gravitational action is the standard Einstein-Hilbert term:
\be
{\cal L} =  \frac{1}{16\pi G_{\rm N}} \sqrt{-g} R + {\cal L}_{\rm DM}[g_{\mu\nu}] + {\cal L}_{\rm b}[\tilde{g}_{\mu\nu}]\,.
\label{introL}
\ee
Dark matter, described by ${\cal L}_{\rm DM}$, couples to $g_{\mu\nu}$. Ordinary matter (`baryons'), described by ${\cal L}_{\rm b}$, couples
to a different metric $\tilde{g}_{\mu\nu}$ to be defined shortly.

Let us first discuss the DM component. We will be primarily interested in cosmological observables on linear scales, which are determined by the
expansion and linear growth histories. On those scales, all we really know about the dark matter is that it behaves to a good approximation
as a pressureless perfect fluid. Thus, to remain agnostic about the DM microphysics, we shall treat DM in the hydrodynamical limit as a perfect fluid. This description of course breaks down on non-linear scales, where the microphysical nature of DM becomes important. However, as we will see it is straightforward to ``complete" our fluid model
with any microphysical theory of DM, be it WIMPs, axions, Bose-Einstein Condensate, superfluid {\it etc.} In other words the fluid approximation is 
made for simplicity, not out of necessity. 

We therefore treat DM within the effective field theory description of perfect fluids \cite{Dubovsky:2005xd, Dubovsky:2011sj,  Endlich:2010hf,Wang:2013ic}.  At low energy, a fluid is described by three Lorentz scalars $\phi^I(x^\mu)$, $I = 1,2,3$, specifying the comoving position of each fluid element as a function of laboratory space-time coordinates $x^\mu$. The ground state configuration is $\phi^I = x^I$, while small perturbations above this state describe phonon excitations. In the absence of vorticity, the description simplifies to a single degree of freedom $\Theta$, corresponding to the longitudinal degree of freedom responsible for laminar flow.  This truncation is consistent at the classical level, thanks to Kelvin's theorem. In the presence of the direct coupling between DM and baryons, two fluid descriptions strictly speaking are not equivalent, see the Appendix for details.
However, since we are interested in the laminar cosmological evolution of DM, the simplified description in terms of a single scalar field will suffice. Specifically, the large scale evolution of dark matter can be conveniently described by
\be
{\cal L}_{\rm DM}=\sqrt{-g}P(X)\;,\quad X=-g^{\mu\nu}\pd_\mu\Theta \pd_\nu \Theta \;.
\ee
Here we take $\Theta$ to have mass dimension $[\Theta]=M^{-1}$ and the function $P(X)$ to have $[P]=M^4$.
The stress tensor of the action ${\cal L}_{\rm DM}$ is given by
\be
T_{\mu\nu} = 2P_{,X}\partial_\mu\Theta\partial_\nu\Theta + Pg_{\mu\nu} \,.
\label{TmunuDM}
\ee
This matches to the perfect fluid form $T_{\mu\nu}=(\rho_{\rm DM}+P_{\rm DM})u_\mu u_{\nu}+P_{\rm DM}g_{\mu\nu}$, with the identification
\begin{align}
\rho_{\rm DM} =2P_{,X}(X)X-P(X)\;,\quad P_{\rm DM}= P(X)\;,\quad u_\mu =-\frac{1}{\sqrt{X}}\pd_\mu\Theta\;.
\label{Duality1}
\end{align}
For our analysis we will not need to specialize to a $P(X)$. All we need is for $P(X)$ to describe non-relativistic particles, such that $P_{\rm DM}\ll \rho_{\rm DM}$.
This amounts to $XP_{,X} \gg P$, in which case
\be
\rho_{\rm DM} \simeq 2P_{,X}X\,.
\label{rhoDMNR}
\ee

\subsection{Baryon action}

Baryons couple to the metric $\tilde{g}_{\mu\nu}$, constructed from $g_{\mu\nu}$ and the various parameters of the DM component: in the hydrodynamical limit, these are the DM density, pressure, 4-velocity, bulk viscosity and shear viscosity. However, since the DM fluid is treated as approximately perfect and assumed nearly pressureless, the only quantities at our disposal are the 4-velocity $u_\mu$ and density, or equivalently $X$. Therefore, the most general form for $\tilde{g}_{\mu\nu}$ is 
\be
\tilde{g}_{\mu\nu}=-Q^2(X)u_\mu u_\nu + R^2(X)\left( g_{\mu\nu}+u_\mu u_\nu \right)\,,
\label{jordanmetric}
\ee
where $R$ and $Q$ are thus far arbitrary functions. The tensor $g_{\mu\nu}+u_\mu u_\nu$ is recognized as the 3-metric orthogonal to the DM velocity. 
The inverse metric is
\be
\tilde{g}^{\mu\nu}=-Q^{-2}(X)u^\mu u^\nu + R^{-2}(X)\left( g^{\mu\nu}+u^\mu u^\nu \right) \,.
\ee
The determinants are related by $\sqrt{-\tilde{g}} = QR^3\sqrt{-g}$. Equivalently, the metric~\eqref{jordanmetric} can be expressed as 
\be
\tilde{g}_{\mu\nu}=R^2(X) g_{\mu\nu} + S(X) \partial_\mu \Theta \partial_\nu\Theta\,,
\label{jordanmetric2}
\ee
where we have introduced
\be
S(X) \equiv \frac{R^2(X)-Q^2(X)}{X}\,.
\label{Sdef}
\ee
This latter form will be helpful when varying the action to obtain the Einstein field equations.

\subsection{Equations of Motion}

Our action~\eqref{introL} is given by
\be
{\cal L} =  \frac{1}{16\pi G_{\rm N}} \sqrt{-g} R + \sqrt{-g}P(X)+ {\cal L}_{\rm b}[\tilde{g}_{\mu\nu}]\,,
\label{theory}
\ee
with $\tilde{g}_{\mu\nu}$ given in \eqref{jordanmetric}. The equation of motion for DM can be obtained by taking the functional derivative of the action with respect to the dark matter field $\Theta$. Explicitly it is given by
\be 
\pd_\nu \Bigg( \bigg[\left(
2P_{,X}+Q R^{3}\tilde{T}^{\alpha\beta}_{\rm b}(2 R R_{,X} g_{\alpha \beta}+S_{,X} \pd_\alpha\Theta \pd_\beta \Theta)
\right)g^{\mu\nu}-Q R^3 S\, \tilde{T}^{\mu\nu}_{\rm b}  \bigg]\sqrt{-g}\pd_\mu \Theta 
\Bigg)=0\;,
\label{DM EoM pressureless baryon}
\ee
where $\tilde{T}_{\rm b}^{\mu\nu}$ is the Jordan-frame energy-momentum tensor for baryons,
\be
\tilde{T}_{\rm b}^{\mu\nu} = \frac{2}{\sqrt{-\tilde{g}}}\frac{\delta{\cal L}_{\rm b}}{\delta \tilde{g}_{\mu\nu}}\,.
\ee
The equation of motion for baryons sector follows from the conservation equation for this stress-tensor:\footnote{ The conservation equation in the Jordan frame follows from the fact that the baryon action $\int\!{\rm d}^4 x{\cal L}_b(\tilde g_{\mu\nu})$ is invariant under coordinate transformations, see e.g.~\cite{Weinberg:1972kfs}. }
\be
\label{conser TJ}
\tilde{\na}_\mu \tilde{T}^{\mu\nu}_{\rm b}=0 \,.
\ee

The Einstein field equations are obtained as usual by varying the action with respect to $g_{\mu\nu}$. For this purpose it is useful to note the relation between the variations of the two metrics:
\be
\delta\tilde{g}_{\mu\nu} = R^2\delta g_{\mu\nu} + \left(2RR_{,X}g_{\mu\nu} + S_{,X}\partial_\mu\Theta\partial_\nu\Theta\right)g^{\alpha\kappa}g^{\beta\lambda}\partial_\alpha\Theta\partial_\beta\Theta \,\delta g_{\kappa\lambda}\,.
\ee
The result is
\be
G_{\mu\nu} = 8\pi G_{\rm N} \Bigg[ T_{\mu\nu} + QR^3 \tilde{T}_{\rm b}^{\kappa\lambda} \bigg(R^2 g_{\kappa\mu}g_{\lambda\nu} + \Big(2RR_{,X}g_{\kappa\lambda} + S_{,X} \partial_\kappa\Theta\partial_\lambda\Theta\Big)\partial_\mu\Theta\partial_\nu\Theta \bigg)\Bigg] \,,
\label{Einsteineqns}
\ee
where the DM stress-energy tensor $T_{\mu\nu}$ was given in~\eqref{TmunuDM}. 

\section{Overview of the Mechanism}
\label{overviewsection}

Before diving into a detailed description, it is worth giving a simplified overview of the mechanism we have in mind. 
On a spatially-flat cosmological background, ${\rm d}s^2 = -{\rm d}t^2 + a^2(t) {\rm d}\vec{x}^2$, the cosmological densities
are functions of the scale factor. In other words, in this case \eqref{jordanmetric} reduces to
\be
\tilde{g}_{\mu\nu}={\rm diag}\left( -Q^2(a),R^2(a)a^2,R^2(a)a^2,R^2(a)a^2 \right)\,.
\label{genmetric}
\ee
There is much freedom in specifying the functions $R$ and $Q$. In general it must satisfy two conditions. First, to ensure that gravity is
standard in the early universe, the coupling must become trivial in the limit of high DM density: $R,~Q  \rightarrow {\rm constant}$.
By rescaling coordinates we can set the constant to unity without loss of generality, hence
\be
R,~Q \rightarrow 1~~~{\rm as}~~~ \rho_{\rm DM}\rightarrow \infty\,.
\label{screening}
\ee
Switching for a moment to a microphysical description, with $\rho_{\rm DM} \rightarrow m\bar{\psi}\psi$ (for fermionic DM) or $m^2\phi^2$ (for bosonic DM), 
this condition also ensures that gravity is standard in high density regions in the present universe, such as galactic halos. Thus~\eqref{screening} enforces a {\it screening 
mechanism} of a remarkably simple kind --- unlike other screening mechanisms~\cite{Joyce:2014kja}, which generally involve solving
the intricate non-linear dynamics of a scalar field, here the deviations from standard gravity are directly determined by 
the (suitably coarse-grained) local DM density.

The second condition is that $R$ and $Q$ should behave at late times in such a way that baryons experience accelerated expansion.\footnote{We mention in passing that the future asymptotic behavior of $R$ and $Q$ (as $\rho_{\rm DM}\rightarrow 0$) is of course not constrained by observations. One could for instance impose that the coupling function once again becomes trivial in this limit: $R,~Q \rightarrow {\rm constant}$ as $\rho_{\rm DM}\rightarrow 0$, where the constant is larger than unity. In this case the present phase of cosmic acceleration would be a transient phenomenon.} Although the Einstein-frame scale factor is always decelerating, the expansion history inferred by baryons is governed by the physical scale factor 
\be
\tilde{a} = R a\,.
\ee
which will be accelerating if $R$ grows sufficiently fast at the present time. In fact for suitable $R$ and $Q$ this can exactly match the $\Lambda$CDM
expansion history. For future purposes, it will be convenient to define a ``rate function",
\be
f  \equiv  \frac{{\rm d}\ln a}{{\rm d}\ln \tilde{a}} = 1 - \frac{{\rm d}\ln R}{{\rm d}\ln\tilde{a}}\,,
\label{fdef}
\ee
whose physical meaning will become clear shortly. Since $R$ increases with time, we see that $f \leq 1$. Furthermore, the Hubble parameters in the two frames are related simply by
\be
\tilde{H}=\frac{H}{Q}\frac{{\rm d ln }\,\tilde{a}}{{\rm d ln }\,a}=\frac{H}{fQ}\,.
\label{Hubbles}
\ee

A comment about the Einstein-frame expansion history. Clearly, in the approximation that one ignores the backreaction of baryons, the Einstein-frame scale factor describes standard Einstein-de Sitter evolution:
\be
a(t) \sim t^{2/3}\,.
\ee
Remarkably, as we will see this result remains true when including the contribution of baryons, for {\it any} choice of $Q(a)$ and $R(a)$, to the extent that DM and baryons are 
separately pressureless fluids. 

What about the growth of density perturbations? Since DM experiences Einstein-de Sitter expansion, density perturbations in the linear regime grow as usual proportional to the scale factor,
\be
\delta \equiv \frac{\delta\rho_{\rm DM}}{\rho_{\rm DM}} \sim a\,.
\ee
This is at first sight worrisome, since a key role played by dark energy is to slow down the growth of structures, consistent with observations. 
However in our model the {\it observed} growth rate should be measured relative to the {\it physical} scale factor $\tilde{a}$.
Following~\cite{Linder:2005in} we define a rescaled growth factor as
\be
g  \equiv \frac{\delta}{\delta_i \tilde{a}} = \frac{a}{\tilde{a}}=\frac{1}{R}\,.
\label{gdef}
\ee
Here $\delta_i$ is the initial density perturbation.
In other words, the rescaled growth factor increases at late times, $g$ 
becomes less than unity, as desired. Similarly, the observed growth rate is
\be
\frac{{\rm d}\ln \delta}{{\rm d}\ln \tilde{a}}  = \frac{{\rm d}\ln \delta}{{\rm d}\ln a} \frac{{\rm d}\ln a}{{\rm d}\ln \tilde{a}} =  \frac{{\rm d}\ln a}{{\rm d}\ln \tilde{a}} = f\,.
\ee
This nicely matches the function $f$ introduced in~\eqref{fdef}. Thus the growth rate is less than unity at late times, as if there were dark energy. 

Physically speaking, this can be understood as a time dilation effect. Although perturbations grow unimpeded in  the Einstein frame, the Einstein-frame universe is younger than the `Jordan-frame' universe experienced by baryons. Hence, from the perspective of ordinary matter, cosmic structures appear less developed than in a pure CDM universe, as if there were dark energy.  We will come back in Sec.~\ref{cosmoobs} to a more quantitative analysis of cosmological observables.

\section{Background Cosmology}
\label{Backgroundandperturb}

We specialize the equations of motion~\eqref{DM EoM pressureless baryon}$-$\eqref{Einsteineqns} to a cosmological background, ${\rm d}s^2 = -{\rm d}t^2 + a^2(t){\rm d}\vec{x}^2$.
For this purpose, we will remain general and not assume anything about the DM and baryonic equations of state until Sec.~\ref{P=0limit}, where we will specialize the results to the 
physically-relevant case of non-relativistic matter.  

\subsection{Background expansion: general results}

By symmetry the DM field depends only on time, $\Theta = \Theta(t)$, such that $X = \dot{\Theta}^2(t)$. The physical metric is given by
\be
\tg_{\mu\nu}=\text{diag} \left\{-Q^2 \;,R^2 a^2(t)\;,R^2 a^2(t)\;,R^2 a^2(t) \right\}\;.
\ee
As usual the baryon component can be treated as a perfect fluid, with
\be
\label{baryon J ST}
\tilde{T}^{\mu\nu}_{\rm b}=\left(\tilde{\rho}_{\rm b}+\tilde{P}_{\rm b}\right){\tilde u}_{\rm b}^\mu{\tilde u}_{\rm b}^\nu+\tilde{P}_{\rm b} \tilde{g}^{\mu\nu}\;,
\ee
where the baryon fluid 4-velocity ${\tilde u}_{\rm b}^\mu$ is unit time-like with respect to $\tilde{g}_{\mu\nu}$:
\be
\tilde{g}_{\mu\nu} {\tilde u}_{\rm b}^\mu{\tilde u}_{\rm b}^\nu = -1\,. 
\ee
The conservation equation~\eqref{conser TJ} implies 
\be
\frac{{\rm d}\tilde{\rho}_{\rm b}}{{\rm d}\ln \tilde{a}} = - 3 \left(\tilde{\rho}_{\rm b}+\tilde{P}_{\rm b}\right)\,,
\label{baryonconserv}
\ee
where $\tilde{a} = Ra$. In particular, if the baryons have negligible pressure, then $\tilde{\rho}_{\rm b} \sim \frac{1}{\tilde{a}^3} = \frac{1}{R^3a^3}$.
To keep the treatment general in what follows we will allow for arbitrary baryon equation of state.

Meanwhile, the DM equation~\eqref{DM EoM pressureless baryon} reduces to
\be
\frac{{\rm d}}{{\rm d}t}  \Bigg( \left[ -P_{,X} + QR^3\left(\frac{Q_{,X}}{Q} \tilde{\rho}_{\rm b} - 3\frac{R_{,X}}{R} \tilde{P}_{\rm b}\right)\right] a^3\dot{\Theta}\Bigg) = 0\,,
\label{DMeom}
\ee
where we have used $g_{\kappa\lambda}\tilde{T}^{\kappa\lambda}_{\rm b} = - Q^{-2}\tilde{\rho}_{\rm b} + 3 R^{-2} \tilde{P}_{\rm b}$. 
Without loss of generality, we can assume $\dot{\Theta} > 0$, hence for the background $\dot{\Theta} = \sqrt{X}$. Then, recalling from~\eqref{Duality1}
that $\rho_{\rm DM} = 2P_{,X}X - P$, the above can be integrated to give
\be
\rho_{\rm DM} = \Lambda_{\rm DM}^4\sqrt{\frac{X}{X_{\rm eq}}} \left(\frac{a_{\rm eq}}{a}\right)^3 - P + 2X QR^3 \left(\frac{Q_{,X}}{Q} \tilde{\rho}_{\rm b} - 3\frac{R_{,X}}{R} \tilde{P}_{\rm b}\right) \,,
\label{rhoDMsoln}
\ee
where the `eq' subscript indicates matter-radiation equality. Since by assumption $Q\simeq R \simeq 1$ in the early universe, and since $P$ will soon be assumed negligible for non-relativistic DM, $\Lambda_{\rm DM}^4$ will be identified as the DM mass density at equality. 

The $(0,0)$ component of the Einstein equations~\eqref{Einsteineqns} yields the Friedmann equation:
\be
3H^2 = 8\pi G_{\rm N}  \left( \rho_{\rm DM} + \rho_{\rm b} \right) \,,
\ee
where we have defined an effective, Einstein-frame baryon density:
\be
\rho_{\rm b} \equiv QR^3 \left(\tilde{\rho}_{\rm b} \left( 1 - 2X\frac{Q_{,X}}{Q}\right) + 6X \frac{R_{,X}}{R} \tilde{P}_{\rm b}\right) \,.
\ee
Substituting~\eqref{rhoDMsoln}, the Friedmann equation becomes
\be
3H^2 = 8\pi G_{\rm N} \left[\Lambda_{\rm DM}^4 \sqrt{\frac{X}{X_{\rm eq}}}\left(\frac{a_{\rm eq}}{a}\right)^3 - P + QR^3\tilde{\rho}_{\rm b} \right]\,.
\label{friedmann}
\ee

The $(i,j)$ components, on the other hand, give the `pressure' equation:
\be
2\frac{\ddot{a}}{a} + \frac{\dot{a}^2}{a^2} = - 8\pi G_{\rm N} \left(P + P_{\rm b} \right)\,,
\label{adoubledot}
\ee
where the effective baryon pressure is 
\be
P_{\rm b} \equiv Q R^3 \tilde{P}_{\rm b}\,.
\ee
In particular, if the baryons are pressureless with respect to the physical metric, they are also pressureless with respect to the Einstein-frame metric. 
Finally, as a check it can easily be verified that any two of the DM equation of motion~\eqref{DMeom}, Friedmann equation~\eqref{friedmann}
and pressure equation~\eqref{adoubledot} imply the third, as it should.  

\subsection{Specializing to pressureless components}
\label{P=0limit}

The above equations simplify tremendously when specialized to the physically-relevant case of nearly pressureless matter components: 
\be
\tilde{P}_{\rm b} \simeq 0 \,;\qquad P \ll 2X P_{,X}\,.
\label{smallpressure}
\ee
As noted before, for baryons the conservation equation~\eqref{baryonconserv} in this case implies $\tilde{\rho}_{\rm b} \sim \tilde{a}^{-3}$, {\it i.e.},
\be
\tilde{\rho}_{\rm b} = \frac{\Lambda_{\rm b}^4}{R^3} \left(\frac{a_{\rm eq}}{a}\right)^3 \,.
\label{rhobsoln}
\ee
Since $R\simeq 1$ in the early universe, $\Lambda_{\rm b}^4$ is identified with the baryon mass density at equality.

More importantly, since the right-hand side of the `pressure' equation~\eqref{adoubledot} vanishes in this limit, 
the background is {\it identically matter-dominated}
\be
a(t) \sim t^{2/3}\,.
\ee
This result holds from matter-radiation equality all the way to the present time, irrespective of the coupling between the two species.
In particular, the total energy density that can be read off from~\eqref{friedmann},
\be\label{rhotot}
\rho_{\rm tot} \equiv \frac{3H^2}{8\pi G_{\rm N}} \simeq \left[\Lambda_{\rm DM}^4 \sqrt{\frac{X}{X_{\rm eq}}} + Q(X) \Lambda_{\rm b}^4\right] \left(\frac{a_{\rm eq}}{a}\right)^3\,,
\ee
where we have neglected the $P$ term and substituted~\eqref{rhobsoln}, redshifts exactly as dust,
\be
\rho_{\rm tot} \sim \frac{1}{a^3}\,.
\ee
It is worth stressing that, remarkably, this result holds for any choice of $Q(X)$ and $R(X)$! This means that the dynamical equation dictates the combination in the square bracket of \eqref{rhotot} to be time independent, for any $Q(X)$.

\section{Linear Perturbations and Stability}
\label{linpert}

In this Section we study the stability of linear perturbations about the cosmological background. To simplify the analysis, we focus on modes that are well-inside the horizon, 
such that mixing with gravity can be ignored. In this regime, the Einstein-frame metric can be treated as a flat, unperturbed metric. 

Our criteria for stability include the usual requirements that the kinetic and gradient matrices be positive-definite. But there is more. Of the two propagating scalar modes,
we will find that one mode propagates with a sound speed that vanishes identically, as a result of the baryon fluid being pressureless. The expression for the sound speed of the second
mode, however, is more complicated and depends explicitly on the form of DM-baryon interactions. In the absence of interactions, it reduces to the DM sound speed, which is arbitrarily small given our assumption of nearly pressureless DM. The presence of interactions, however, generically modifies the result and can give rise to relativistic sound speed. In particular, this is unavoidable in the case of conformal coupling $Q(X) = R(X)$.

A relativistic sound speed is undesirable, for it can give rise to unwanted oscillatory features in the matter power spectrum~\cite{Sandvik:2002jz,Wang:2013qy}. As we will show in Sec.~\ref{confcouplingruledout}, this seems unavoidable in the particular case of conformal coupling $Q(X) = R(X)$. However, the key point is that this conclusion is special to the conformal limit. More general, disformal couplings (with $Q \neq R$) do allow stable perturbations with arbitrarily small sound speeds. We will demonstrate this emphatically in Sec.~\ref{pressurecase} with the `maximally disformal', or `pressure' coupling, corresponding to $Q = 1$. 

\subsection{General demixing}

For simplicity, we once again model the baryon component as pressureless, $\tilde{P}_{\rm b} = 0$. Since the Einstein-frame metric is approximated as flat in this analysis, the background baryon physical density~\eqref{rhobsoln} is approximately constant. Without loss of generality we can set the scale factor at the time of interest to unity: $a_*=a(t_*)=1$.
Similarly the DM conservation~\eqref{DMeom} tells us that the background value of the DM field can be treated as time independent, --- i.e. ${\bar X} = X(t_*) = {\rm const.}$ in our approximation, --- and hence that ${\bar \rho}_{\rm DM} \simeq 2 \bar{X} P_{,X}({\bar X}) = {\rm const.}$ as well.
By a trivial rescaling of $\Lambda_{\rm b}$, we will write the background baryon density at the time of interest as 
\be
\overline{\tilde{\rho}}_{\rm b} = \frac{\Lambda_{\rm b}^4}{\bar{R}^3} \,.
\label{rhobsoln2}
\ee

We perturb the DM field as $\Theta = \bar{\Theta}(t) + \theta(t,\vec{x})$, such that $X = \bar{X} + 2\dot{\bar{\Theta}}\dot{\theta}$ at linear order. Dropping bars to simplify notation,
the linearized perturbation of the physical metric is given by
\begin{align}
\delta\tg_{\mu\nu} = \Big(2R_{,X} R g_{\mu\nu} + S_{,X} X \delta^0_{~\mu} \delta^0_{~\nu}\Big)2 \sqrt{X}\dot{\theta} + 
2 S\sqrt{X} \delta^0_{~(\mu}\partial_{\nu)}\theta \,,
\end{align}
where, in anticipation of ignoring the mixing with gravity, we have fixed the Einstein-frame metric to its unperturbed, FRW form $g_{\mu\nu} = {\rm diag}\left( -Q_*^2,R_*^2,R_*^2,R_*^2 \right)$. The baryon variables are the density perturbation ${\tilde \delta}_{\rm b}\equiv\frac{\delta{\tilde \rho}_{\rm b}}{{\tilde \rho}_{\rm b}}=\Lambda_{\rm b}^{-4}\delta{\tilde \rho}_{\rm b} R^3$ and velocity perturbation $u_i^{\rm b}$. 

The linearized DM equation reads
\begin{align}
&\left(-2P_{,X}-4 X P_{,XX}+2\Lambda_{\rm b}^4\left(Q_{,X}+2X \bQ_{,XX}+\frac{6X Q_{,X}R_{,X}}{R}\right)
\right) \ddot \theta  \nn \\
& +
\left[
2P_{,X}- \Lambda_{\rm b}^4\left(\frac{S}{Q} + 2Q_{,X}\right)
\right]\vec{\nabla}^2 \theta 
+2\Lambda_{\rm b}^4 \sqrt{X}Q_{,X} \dot{\tilde \delta}_{\rm b}
-\frac{\Lambda_{\rm b}^4 \sqrt{X}S}{R^2}\pd_i u_i^{\rm b}=0\;.
\label{DMlinear}
\end{align}
Meanwhile, the energy and momentum conservation equations for baryons reduce to
\bea
\nn
& & \dot{\tilde \delta}_{\rm b} +6\sqrt{X} \frac{R_{,X}}{R}\ddot \theta+\frac{Q}{R^2}\pd_i u_i^{\rm b}=0\;,\\
& & \pd_t u_i^{\rm b} +  \sqrt{X}\left(2Q_{,X}+\frac{S}{Q}\right) \pd_i{\dot \theta} =0\;.
\label{baryonlinear}
\eea
Thus we have a system of three coupled partial differential equations for three variables $\theta$, $\tilde \delta_{\rm b}$ and $u_i^{\rm b}$. 
By solving the first of~\eqref{baryonlinear} for $\dot{\tilde \delta}_{\rm b}$ and substituting the result into~\eqref{DMlinear}, the problem is reduced
to two coupled equations for  $\theta$ and $u_i^{\rm b}$. Focusing on the longitudinal mode $u_i^{\rm b}=\pd_i u^{\rm b}$, and working in Fourier space,
these two equations combine into a matrix equation:
\begin{align}
\left( \begin{array}{cc}
-A_{11}\omega^2-B_{11} k^2 & A_{12} k^2\\
A_{12}\omega k & A_{22} \omega k
\end{array} \right)
\left( \begin{array}{c}
\theta\\
u^{\rm b}
\end{array} \right)=0\;,
\end{align}
where 
\begin{align}
A_{11}&=-2 P_{,X}-4X P_{,XX}+2\Lambda_{\rm b}^4\left(Q_{,X}+2X Q_{,XX}\right)\,;\nn\\
B_{11}&=2P_{,X}-\frac{\Lambda_{\rm b}^4}{Q}(S+2Q_{,X}Q)\,;\nn\\
A_{12}&=\Lambda_{\rm b}^4 \frac{\sqrt{X}}{R^2}(S+2Q_{,X}Q)\,;\nn\\
A_{22}&=\Lambda_{\rm b}^4\frac{Q}{R^2}\,.
\end{align}
Diagonalizing this matrix, it follows immediately that the dispersion relations for the decoupled modes are
\begin{align}
\omega=0\;;\quad \text{and}~~~\omega^2=\frac{P_{,X}-\frac{\Lambda_{\rm b}^4}{2R^2}(Q-2X Q_{,X})(S+2Q Q_{,X})}{P_{,X}+2 X P_{,XX}-\Lambda_{\rm b}^4\left(Q_{,X}+2X Q_{,XX}\right)}\,k^2\;.
\label{dispersion}
\end{align}

Thus we see that one mode propagates with zero sound speed, irrespective of the choice of coupling functions. The vanishing of $c_s$ in this case traces
back to our pressureless assumption for the baryon component; indeed, it is straightforward to show that allowing for $\tilde{P}_{\rm b}\neq 0$ would make $c_s$
non-zero, though the expression is fairly complicated and therefore not particularly illuminating. 

Our main interest, however, is in the second mode. At sufficiently early times, when $R \simeq Q \simeq 1$ and the components are decoupled, the dispersion relation reduces to the standard expression for $c_s$ derived from $P(X)$. At late times, however, when cosmic acceleration kicks in and the functions $R(X)$ and $Q(X)$ grow by order unity, there is no reason {\it a priori} for the sound speed to remain small (or even real, for that matter). Instead, these functions must be selected such that $0<c_s^2\ll 1$, as desired. Below we consider two special cases for the coupling: the conformal case, corresponding $Q(X)=R(X)$, and the `maximally disformal' or `pressure' coupling, corresponding to $Q = 1$. 

\subsection{Conformal Coupling}
\label{confcouplingruledout}

The conformal coupling $Q(X)=R(X)$, being the simplest, deserves separate consideration. In this case $S = 0$, and the non-zero sound speed read off from~\eqref{dispersion} reduces to
\begin{align}
c_s^2=\frac{P_{,X}-\Lambda_{\rm b}^4 Q_{,X}+2\Lambda^4_{\rm b} X \frac{Q_{,X}^2}{Q}}{P_{,X}-\Lambda_{\rm b}^4 Q_{,X}+2X (P_{,XX}-\Lambda_{\rm b}^4 Q_{,XX})}\;.
\end{align}
For stability we need both numerator and denominator to be positive. Furthermore we require that $0 < c_s^2 \ll 1$. These conditions amount to 
\bea
\nonumber
P_{,X} &>& \Lambda_{\rm b}^4 Q_{,X}\left(1 - 2X \frac{Q_{,X}}{Q}\right)\,;\\
P_{,XX} &\gg & \Lambda_{\rm b}^4Q \left( \frac{Q_{,X}^2}{Q^2}+\frac{Q_{,XX}}{Q} \right) \,.
\label{csineq}
\eea

It is straightforward to argue, however, that these conditions are incompatible with our goal of achieving cosmic acceleration. To see this, recall that $P(X)$ is chosen such that, in the absence of coupling to baryon, the would-be DM sound speed,
\be
c_{\rm DM}^2 \equiv \frac{P_{,X}}{P_{,X} - 2XP_{,XX}}\,,
\ee
is always small. This is an intrinsic property of $P(X)$ that remains true even when we turn on the coupling of baryons, the only difference being that $c_{\rm DM}^2$ no longer represents the propagation speed of physical modes. Nevertheless, it is useful to cast the argument below in terms of $c_{\rm DM}^2\ll 1$. 

For starters, we note that by definition  
\be
c_{\rm DM}^2 = \frac{{\rm d}P}{{\rm d}\rho_{\rm DM}}  = \frac{P_{,X}}{\rho_{\rm DM}} \frac{{\rm d}\ln a}{{\rm d}\ln \rho_{\rm DM}} \frac{{\rm d}X}{{\rm d}\ln a} \simeq \frac{1}{2} \frac{{\rm d}\ln a}{{\rm d}\ln \rho_{\rm DM}}\frac{{\rm d}\ln X}{{\rm d}\ln a}\,,
\label{cDMdef}
\ee
where in the last step we have used $\rho_{\rm DM} \simeq 2 XP_{,X}$. Although DM-baryon interactions alter the usual scaling $\rho_{\rm DM}\sim 1/a^3$ --- see~\eqref{rhoDMsoln} --- it is nevertheless reasonable to assume that $\left\vert\frac{{\rm d}\ln \rho_{\rm DM}}{{\rm d}\ln a}\right\vert \sim {\cal O}(1)$ to obtain a reasonable cosmology. In that case we learn that
\be
\left\vert \frac{{\rm d}\ln X}{{\rm d}\ln a}\right\vert \sim c_{\rm DM}^2 \ll 1\,, 
\label{Xvary}
\ee
in other words $X(a)$ is almost constant. Furthermore, in order to mimic cosmic acceleration for the physical scale factor $\tilde{a} = Q\, a$, clearly it is necessary that $\frac{{\rm d}\ln Q}{{\rm d}\ln a} \sim {\cal O}(1)$ at late times. Combining this with~\eqref{Xvary}, we obtain
\be
\left\vert \frac{{\rm d}\ln Q}{{\rm d}\ln X} \right\vert \sim \frac{1}{c_{\rm DM}^2} \gg 1\,.
\label{Qvary}
\ee
Thus, barring any cancellation the second of our desired inequalities~\eqref{csineq} amounts to
\be
X^2 P_{,XX} \gg  \frac{\Lambda_{\rm b}^4Q}{c_{\rm DM}^4}\,.
\label{csineq2}
\ee
Meanwhile, from the definition~\eqref{cDMdef} of $c_{\rm DM}^2$ it is easy to see that, in the limit $c_{\rm DM}^2 \ll 1$, we have 
\be
X^2P_{,XX} \simeq \frac{XP_{,X}}{2c_{\rm DM}^2}\simeq \frac{\rho_{\rm DM}}{4c_{\rm DM}^2}\,, 
\label{intermresult}
\ee
where we have used~\eqref{rhoDMNR}. Furthermore, using the relation~\eqref{rhobsoln2} with $R = Q$, the inequality~\eqref{csineq2} simplifies to
\be
\rho_{\rm DM} \gg Q^4 \frac{\tilde{\rho}_{\rm b}}{c_{\rm DM}^2} > \frac{\tilde{\rho}_{\rm b}}{c_{\rm DM}^2} \,.
\ee
Here it is understood that ${\tilde \rho}_{\rm b}$ and $\rho_{\rm DM}$ are the baryon and DM energy density at the time of interest.
It is easy to see that the above inequality is impossible to satisfy. For any reasonable cosmology we expect $\rho_{\rm DM}$ and $\tilde{\rho_{\rm b}}$ to differ by at most an order of magnitude (see \eqref{rhoDMsoln}), which
is clearly insufficient to overcome the factor of $1/c_{\rm DM}^2$ on the right hand side.

What we have learned is that, for conformal coupling, the stability/phenomenological requirement $0 < c_s^2 \ll 1$ embodied in~\eqref{csineq} does not allow $Q$ to vary sufficiently to drive late-time cosmic acceleration. Instead $Q$ must remain approximately constant. Possible loopholes in this argument are that $Q$ may be fine-tuned to keep the right-hand side of~\eqref{csineq} artificially small and/or $\left\vert\frac{{\rm d}\ln \rho_{\rm DM}}{{\rm d}\ln a}\right\vert$ artificially large. Although there may exist special functional forms for $Q(X)$ for which this is the case, we will not pursue the conformal case further. Instead we turn to the more promising case of disformal coupling, $Q\neq R$.


\subsection{Maximally-Disformal Coupling}
\label{pressurecase}

It should be clear from the dispersion relation in~\eqref{dispersion} that the requirement $c_s\ll 1$ forces $Q(X)$ to be a slowly-varying function. This is what spelled doom for the conformal case --- since everything is controlled by $Q$ in that case, a nearly constant $Q(X)$ implies a negligible impact on the background evolution. In the disformal case $Q\neq R$, however, it is in principle possible to keep $Q$ approximately (or even exactly) constant, while $R(X)$ can have arbitrary time-dependence.

To simplify the discussion, let us focus on the maximally-disformal case where $Q$ remains exactly constant, {\it i.e.}, $Q\equiv 1$ at all times.
In this case~\eqref{dispersion} implies after some manipulation the sound speed:
\be
c_s^2=\frac{1-\frac{\Lambda_{\rm b}^4}{2XP_{,X}}  \left( 1 - \frac{1}{R^2}\right)}{1+\frac{2 X P_{,XX}}{P_{,X}}} \,,
\ee
where we have substituted~\eqref{Sdef} for $S$. Next, using the definition of $c_{\rm DM}^2$, in particular~\eqref{intermresult}, as well as $\rho_{\rm DM} \simeq 2 XP_{,X}$
the sound speed becomes
\be
c_s^2 \simeq c_{\rm DM}^2 \left (1  -  \frac{\Lambda_{\rm b}^4}{\rho_{\rm DM}} \left( 1 - \frac{1}{R^2}\right) \right) = c_{\rm DM}^2 \left (1  -  \frac{ \tilde{\rho}_{\rm b}}{\rho_{\rm DM}} R\left( R^2 - 1\right) \right)\,,
\label{cs2almostfinal}
\ee
where in the last step we have substituted~\eqref{rhobsoln2}. 

Thus in this case the sound speed is proportional to $c_{\rm DM}^2$, which can be made arbitrarily small. It remains to show that $c_s^2$ is also positive. 
To see this, note that~\eqref{rhoDMsoln} with $Q = 1$, $P\ll \rho_{\rm DM}$ and $\tilde{P}_{\rm b} =0$, implies $\rho_{\rm DM} = \Lambda_{\rm DM}^4 \sqrt{\frac{X}{X_{\rm eq}}} \left(\frac{a_{\rm eq}}{a}\right)^3$. Combined with~\eqref{rhobsoln}, we get
\be
\frac{\tilde{\rho}_{\rm b}}{\rho_{\rm DM}} \simeq \frac{\Lambda_{\rm b}^4}{\Lambda_{\rm DM}^4} \sqrt{\frac{X_{\rm eq}}{X}} \frac{1}{R^3}\,.
\ee
Substituting into~\eqref{cs2almostfinal} gives
\be
c_s^2 \simeq c_{\rm DM}^2 \left (1  -  \frac{\Lambda_{\rm b}^4}{\Lambda_{\rm DM}^4}\left(1 - \frac{1}{R^2}\right)  \sqrt{\frac{X_{\rm eq}}{X}}\right)\,.
\ee
By definition $\Lambda_{\rm b}^4$ and $\Lambda_{\rm DM}^4$ set the baryon and DM density at matter-radiation equality, hence $\frac{\Lambda_{\rm b}^4}{\Lambda_{\rm DM}^4}\simeq \frac{1}{6}$. Meanwhile, as argued earlier $X(a)$ is nearly constant in the limit $c_{\rm DM}^2 \ll 1$ --- see~\eqref{Xvary}. Therefore, it follows that $c_s^2$ is positive, as desired. 

\section{Integrating in Fields}
\label{integratein}

Although our scenario does not require any additional degree of freedom beyond DM and ordinary matter, it is sometimes conceptually helpful to ``integrate in"  additional fields to make contact with a language perhaps more familiar to the readers. In case of the conformal coupling, it is sufficient to introduce a massive scalar scalar field $\phi$ for this purpose. In the generic case, on the other hand, the introduction of an additional massive vector field $A_\mu$ is required. Furthermore, for concreteness, and to make contact with particle physics theories of DM, we shall model the DM field as a fermion $\psi$. (The generalization to bosonic DM is of course straightforward.)

\subsection{Scalar-Vector-Tensor Formulation}

Consider the (Einstein-frame) theory 
\bea
\nonumber
{\cal L}  &=& \sqrt{-g} \left(\frac{R}{16\pi G_{\rm N}}  -\frac{1}{2}(\partial\phi)^2  - \frac{1}{2}m_\phi^2 \phi^2 -\frac{1}{4}F_{\mu\nu}^2+\frac{1}{2}m_A^2 A_\mu^2\right) \\
& & -  \sqrt{-g}\left( \left(1 - \frac{\phi}{M}\right) m_\psi\bar{\psi}\psi + \alpha A_\mu\bar{\psi}\gamma^\mu\psi \right)  +\mathcal{L}_{b}[\tilde{g}_{\mu\nu}]\,,
\label{Lscalar}
\eea
where $F_{\mu\nu} = \partial_\mu A_\nu - \partial_\nu A_\mu$ is the field strength for $A_\mu$, $M$ is some arbitrary mass scale, $\alpha$ is a dimensionless constant, and 
\be
\tilde{g}_{\mu\nu}\equiv R^2(\phi)\left( g_{\mu\nu}-\frac{A_\mu A_\nu}{A^2} \right) + Q^2(\phi) \frac{A_\mu A_\nu}{A^2}\,.
\ee
Ignoring the contribution from baryons for a moment, the equations of motion for $\phi$ and $A_\mu$ are 
\bea
\nn
\Box\phi &=& m_\phi^2 \phi  - \frac{m_\psi}{M} \bar{\psi}\psi\,;\\
\nabla_\mu F^{\mu\nu}&=& -m_A^2 A^{\nu}+\alpha \bar{\psi}\gamma^\nu\psi\,.
\label{Aeom}
\eea
For the purpose of solving~\eqref{Aeom} we imagine coarse graining the DM distribution over scales much larger than the interparticle separation, which amounts to a hydrodynamical approximation. In this regime $m_\psi \la \bar{\psi} \psi\ra$ reduces to the DM energy density $\rho_{\rm DM}$, whereas $\la \bar{\psi}\gamma^\mu\psi \ra$ represents the averaged current and is therefore proportional to the $4$-velocity of the fluid element $u^\mu$. Furthermore, assuming that $\phi$ and $A_\mu$ are sufficiently heavy, we can ignore their gradients and treat them as auxiliary fields. Thus, by averaging~\eqref{Aeom} we obtain the expectation value of the auxiliary fields on large ({\it i.e.}, cosmological) scales, with result
\begin{align}
\nn
&\phi=\frac{m_\psi}{Mm_\phi^2}\la \bar{\psi}\psi \ra \simeq \frac{\rho_{\rm DM}}{Mm_\phi^2}\,;\\
&A^\mu=\frac{\alpha}{m_A^2}\la \bar{\psi}\gamma^\mu \psi \ra \sim u^\mu\,.
\label{bgphi}
\end{align}
It is straightforward to convince oneself that this implies the equivalence of~\eqref{Lscalar} and the original action~\eqref{introL}, in the regime that baryons are negligible.

For self-consistency, we should check our approximation of neglecting baryons in the above equations. The baryonic contribution to~\eqref{Aeom} can be readily computed
\begin{align}
\nonumber
\frac{\delta \mathcal{L}_{\rm b}}{\delta \phi}:&\qquad \frac{{\rm d} \mathcal{L}_{\rm b}}{{\rm d} \tilde{g}_{\mu\nu}}\frac{{\rm d} \tilde{g}_{\mu\nu}}{{\rm d}\phi}=\frac{{\rm d} \mathcal{L}_{\rm b}}{{\rm d} \tilde{g}_{\mu\nu}}\left[ 2 R R_{,\phi}(\phi)\left( g_{\mu\nu}-\frac{A_\mu A_\nu}{A^2} \right)+2 Q Q_{,\phi}(\phi)\frac{A_\mu A_\nu}{A^2} \right]\,;\\
\frac{\delta \mathcal{L}_{\rm b}}{\delta A_\alpha}:&\qquad \frac{{\rm d} \mathcal{L}_{\rm b}}{{\rm d} \tilde{g}_{\mu\nu}}\frac{{\rm d} \tilde{g}_{\mu\nu}}{{\rm d}A_\alpha}=-\frac{{\rm d} \mathcal{L}_{\rm b}}{{\rm d} \tilde{g}_{\mu\nu}} \left( R^2-Q^2 \right)\frac{\delta^\alpha_\mu A_\nu+\delta^\alpha_\nu A_\mu-2A_\mu A_\nu A^\alpha/A^2}{A^2}\,.
\end{align}
This expressions can be greatly simplified, if we notice that
\be
\frac{{\rm d} \mathcal{L}_{\rm b}}{{\rm d} \tilde{g}_{\mu\nu}}=\frac{\sqrt{-\tilde{g}}}{2} \tilde{T}_{\rm b}^{\mu \nu}=\frac{\sqrt{-\tilde{g}}}{2}\tilde{\rho}_{\rm b}\tilde{u}_{\rm b}^\mu \tilde{u}_{\rm b}^\nu\,,
\ee
for the pressureless baryon fluid. Moreover, at the background level the velocity $\tilde{u}_{\rm b}^\mu$ is aligned with the dark matter velocity $u^\mu$, and consequently with $A^{\mu}$. After combining everything we get the remarkably simple result
\begin{align}
\nonumber
\frac{\delta \mathcal{L}_{\rm b}}{\delta \phi}:&\qquad \frac{{\rm d} \mathcal{L}_{\rm b}}{{\rm d} \tilde{g}_{\mu\nu}}\frac{{\rm d} \tilde{g}_{\mu\nu}}{{\rm d}\phi}=-R^3Q_{,\phi}(\phi) \tilde{\rho}_{\rm b}\,;\\
\frac{\delta \mathcal{L}_{\rm b}}{\delta A_\alpha}:&\qquad \frac{{\rm d} \mathcal{L}_{\rm b}}{{\rm d} \tilde{g}_{\mu\nu}}\frac{{\rm d} \tilde{g}_{\mu\nu}}{{\rm d}A_\alpha}=0\,.
\end{align}
Therefore, our solution for $A_\mu$ (second of~\eqref{bgphi}), is correct even when including baryons, whereas our solution for $\phi$ (first of~\eqref{bgphi}) 
is accurate to the extent that 
\be
\frac{\rho_{\rm DM}}{M}\gg R^3\left\vert Q_{,\phi}(\phi)\right\vert \tilde{\rho}_{\rm b}\,.
\label{ineqbaryon1}
\ee
This is trivially satisfied for the maximally-disformal coupling, $Q = 1$, telling us that in that case the solution for $\phi$ continues to apply with baryons. 
More generally, it is helpful to cast it in terms of the $X$ variable using the chain rule
\be
Q_{,\phi} =  \frac{{\rm d}\rho_{\rm DM}}{{\rm d}\phi} \frac{{\rm d}P}{{\rm d}\rho_{\rm DM}} \frac{Q_{,X}}{P_{,X}} \simeq 2 M m_\phi^2 c_{\rm DM}^2  \frac{XQ_{,X}}{\rho_{\rm DM}}\,,
\label{Qphichainrule}
\ee
where we have used~\eqref{bgphi}, the definition of $c_{\rm DM}^2 \equiv {\rm d}P/{\rm d}\rho_{\rm DM}$, and $\rho_{\rm DM}\simeq 2 X P_{,X}$. Furthermore, using the fact that $Q,R \geq 1$, our criterion~\eqref{ineqbaryon1} reduces to
\be
\frac{\rho_{\rm DM}}{M^2 m_\phi^2} \gg \frac{\tilde{\rho}_{\rm b}}{\rho_{\rm DM}} c_{\rm DM}^2 \left\vert \frac{{\rm d}\ln Q}{{\rm d}\ln X}\right\vert\,.
\label{ineqbaryon2}
\ee
On the other hand, since $\rho_{\rm DM}$ evolves cosmologically on a Hubble time, then we must require $m_\phi \gg H$ (as well as $m_A \gg H$) in order 
for~\eqref{bgphi} to remain true adiabatically. Hence~\eqref{ineqbaryon2} implies 
\be
\frac{\rho_{\rm DM}}{H^2M^2} \gg \frac{\tilde{\rho}_{\rm b}}{\rho_{\rm DM}} c_{\rm DM}^2 \left\vert \frac{{\rm d}\ln Q}{{\rm d}\ln X}\right\vert\,.
\ee
Even in the conformal case (which, in any case, is undesirable at the level of perturbations as discussed earlier), where $c_{\rm DM}^2 \left\vert \frac{{\rm d}\ln Q}{{\rm d}\ln X}\right\vert \sim {\cal O}(1)$ (see~\eqref{Qvary}), this inequality is easily satisfied by taking $M \ll M_{\rm Pl}$. In the maximally-disformal case $Q = 1$, the bound is of course trivially satisfied.

\subsection{Local Constraints}

The scalar-vector-tensor formulation given by~\eqref{Lscalar} is particularly useful to derive the predicted signals for DM-nucleon scattering (direct detection) and DM annihilation (indirect detection). For this purpose we must determine the field values $\bar{\phi}$ and $\bar{A}_\mu$ assumed in the local environment. Averaging over the local (galactic) DM density, the answer is given by~\eqref{bgphi} (again ignoring baryons). 

We are interested in four-body effective interaction vertex between DM particles and baryons. Expanding about the background values, $\phi = \bar{\phi} + \varphi$ and $A_\mu=\bar{A}_\mu+a_\mu$, the part of the action relevant for local experiments is
\begin{align}
{\cal L} = -\frac{1}{2}(\partial\varphi)^2  - \frac{1}{2}m_\phi^2 \varphi^2 -\frac{1}{4}f_{\mu\nu}^2+\frac{1}{2}m_A^2 a_\mu^2
+ \frac{m_\psi}{M}  \varphi \bar{\psi}\psi -\alpha a_\mu\bar{\psi}\gamma^{\mu}\psi\nonumber \\
+ \frac{Q_{,\phi}(\bar{\phi})}{Q(\bar{\phi})}m_{\rm b}  \varphi\bar{b} b +\frac{Q^2-R^2}{R^2Q^2}m_{\rm b}\frac{a_i}{\bar{A}^0}\bar{b}\gamma^ib+ \ldots
\end{align}
where $m_{\rm b}$ denotes the mass of baryon particle and $f_{\mu\nu}\equiv \partial_\mu a_\nu-\partial_\nu a_\mu$ is the field strength for the vector perturbation. By integrating out $\varphi$ and $a_\mu$ we can write down effective Fermi vertices describing DM-baryon interactions.
This is done as usual by evaluating the DM-baryon scattering amplitude mediated by $\varphi$ and $a_\mu$ exchange. In the limit of large $m_\phi$ and $m_A$, the scalar and vector propagators just
become $1/m_\phi^2$ and $1/m_A^2$ respectively, and the effective Lagrangian reduces to
\be
{\cal L}_{\rm eff} \sim G_{\rm F}^\varphi \bar{\psi}\psi \bar{b} b+G_{\rm F}^a \bar{\psi}\gamma^i\psi \bar{b}\gamma_i b \,.
\ee
The effective Fermi's constants are given by
\begin{align}
\label{scalarfermi}
&G_{\rm F} ^\varphi= \frac{m_\psi m_{\rm b}}{Mm_\phi^2} \frac{Q_{,\phi}(\bar{\phi})}{Q(\bar{\phi})}= 2 \frac{m_\psi m_{\rm b}}{\rho_{\rm DM}}  c_{\rm DM}^2 \frac{{\rm d}\ln Q}{{\rm d}\ln X} \,,\\
&G_{\rm F} ^a=\frac{Q^2-R^2}{R^2Q^2}\frac{m_\psi m_{\rm b}}{\rho_{\rm DM}}\,,
\end{align}
where in the last step of \eqref{scalarfermi} we have substituted~\eqref{Qphichainrule}. 

These effective coupling constants exhibit {\it screening} firstly because they are inversely proportional to $\rho_{\rm DM}$, and secondly because $R\rightarrow Q \rightarrow 1$
at high density. Both factors tend to suppress $G_{\rm F}$ and weaken DM-baryon interactions in regions of high density. Moreover, $G_{\rm F}^\varphi$ is further
suppressed by $c_{\rm DM}^2$, which can be made arbitrarily small. We leave to the future a detailed discussion of direct and indirect detection constraints.
As mentioned earlier our primary interest lies in the maximally-disformal case $Q = 1$, for which $G_{\rm F} ^\varphi$ vanishes.

We would like to finish this section by stressing that we are dealing with the effective field theory. Therefore, we should expect the presence of new degrees of freedom in the spectrum, with mass of the order of the cut-off of the theory. Their presence is required by unitarity, as otherwise the scattering amplitudes would diverge at the cut-off. Therefore, in order to justify the novelty of our scenario, claimed throughout the paper, we need to make sure that these additional degrees of freedom are much heavier than the Hubble scale. At late times the $Q$ and $R$ dependent factors of the Fermi's constants can be approximated as unity, resulting in
\be
G_{\rm F}\sim \frac{m_\psi m_{\rm b}}{\rho_{\rm DM}}\,.
\ee
It is easy to see that the suppression scale of this coupling is greater than $H$, as long as
\be
\rho_{\rm DM}\gg m_\psi m_{\rm b}H^2\,.
\label{cut-off}
\ee
Taking into account that at late times DM density is approximately equal to the total energy density, we can use the Friedmann equation $\rho_{\rm DM}\simeq H^2\mpl^2$ to rewrite \eqref{cut-off} as
\be
\mpl^2\gg m_\psi m_{\rm b}\,.
\ee
Obviously, this inequality is easily satisfied.
In order to give a numerical estimate of the coupling strength let us assume $m_\psi={\rm eV}$, $m_{\rm b}\sim {\rm GeV}$ and $\rho_{\rm DM}\sim {\rm meV}^4$, resulting in
\be
G_{\rm F}\sim \frac{1}{10^{-21}{\rm eV}^2}\,.
\ee
Therefore, the cut-off of our effective theory is many orders of magnitude bigger than the Hubble constant today. This means that all the additional degrees of freedom in our model are much heavier than the curvature scale, manifesting the novelty of our scenario.

\section{Cosmological Observables}
\label{cosmoobs}

In this Section we derive various cosmological observables for our model and compare the results to $\Lambda$CDM predictions. 
For concreteness, we focus on the maximally-disformal case, $Q =1$,  since as discussed in Sec.~\ref{pressurecase} the sound speed
of fluctuations in this case is consistently small and positive. This case is also more predictive, since we are left with a single function $R(X)$ 
to fit against data. For illustrative purposes we will focus on a simple parametrization for this function, involving two parameters, and 
choose parameter values that give a reasonable fit to the data, without attempting a full likelihood analysis to derive a best-fit model.
This is left for future work. One observable we will not consider here is the CMB angular power spectrum, as this requires modifying the CAMB numerical code.
The full derivation of the CMB spectrum will be presented elsewhere~\cite{AdamVin}.

First let us set some conventions. Instead of $R(X)$ it turns out to be convenient to work in terms of its inverse, the rescaled growth factor defined in~\eqref{gdef}:
\be
g  \equiv \frac{a}{\tilde{a}}=\frac{1}{R}\,.
\label{gdef2}
\ee
By rescaling coordinates, we can set the physical scale factor to unity at the present time, $\tilde{a}_0 = 1$, but then in general the present-day Einstein frame scale factor is left unfixed, $a_0 = g_0  \neq 1$. Redshift factors can be defined in both frames as follows:
\be
\tilde{a} = \frac{1}{1 + \tilde{z}}\,;\qquad a = \frac{g_0}{1+z}\,.
\ee
The present time corresponds to $\tilde{z} = z = 0$, as it should. We also recall the rate function $f\equiv  \frac{{\rm d}\ln a}{{\rm d}\ln \tilde{a}}$, introduced in~\eqref{fdef}. This can be expressed in terms of redshift as
\be
f(\tilde{z}) = 1 - (1 + \tilde{z}) \frac{{\rm d}\ln g}{{\rm d}\tilde{z}} \,.
\label{fz}
\ee
We will work in the approximation that baryons are pressureless, $\tilde{P}_{\rm b} = 0$. As discussed in Sec.~\ref{P=0limit}, it follows that the Einstein-frame scale factor behaves exactly as a dust-dominated universe, $a(t) \sim t^{2/3}$, all the way to the present time. In other words,
\be
H(a) = H_0 \left(\frac{a_0}{a}\right)^{3/2}\,.
\ee

\subsection{Fiducial model}
\label{fidmod}

In principle to fix a model we should specify a DM function $P(X)$ and a coupling function $R(X)$, and then solve the DM equation of motion~\eqref{DMeom} to obtain $X(a)$.
Equivalently, we can assume that this has been done already and specify $R(a)$ directly. This gives us an expression for $\tilde{a}(a) = R(a) a$, with which we can calculate various observables. In practice, however, the fitting procedure is simpler if we have an analytic expression for the inverse, $a(\tilde{a})$. While there is a one-to-one correspondence between the latter and the former, this may require numerically solving a transcendental equation. To short-circuit these complications, we will follow an easier route by directly specifying
the function $a(\tilde{a})$, or in other words, $g(\tilde{a})$. This suffices for the purpose of this section, namely to present a proof of principle for the existence of DM-baryon coupling functions that give a reasonable fit to data. 

Specifically we consider as fiducial function the following polynomial form:
\be
a(\tilde{a}) = \tilde{a}+\alpha \tilde{a}^2+\beta \tilde{a}^3\,.
\label{fiducial}
\ee
The corresponding rescaled growth function follows immediately:
\be
g(\tilde{a})=1+\alpha \tilde{a}+\beta \tilde{a}^2\,.
\label{growth}
\ee
The coefficient of the linear term was fixed by the requirement that $a \simeq \tilde{a}$ at early times ({\it i.e.}, for $\tilde{a} \ll 1$).
We have explored the effect of including higher-order terms as well, but this turns out to make little difference in terms of improving the
fit to data. One should keep in mind that this simple functional form is only meant to be valid up to the present time, $\tilde{a}\leq 1$. 
This may be appropriately modified at larger values of $\tilde{a}$, in order to get suitable future asymptotic behavior.

\begin{figure}[t]
\centering
\includegraphics[width=3.5in]{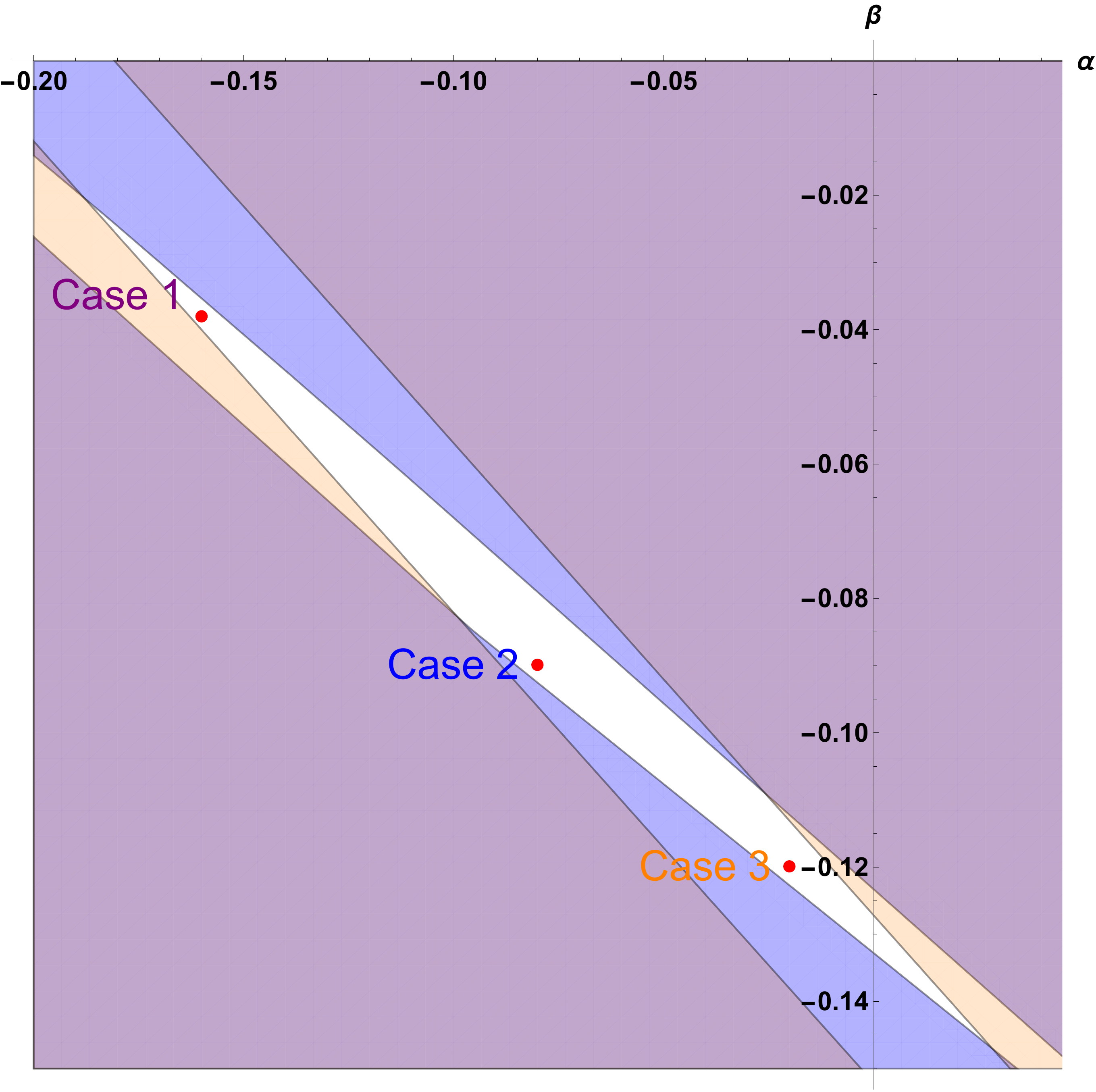}
\caption{\label{alphabetaregion} \small The white region represents the allowed region in $(\alpha,\beta)$ parameter space, where $\alpha$ and $\beta$ are coefficients of our fiducial the polynomial function $a(\tilde{a}) = \tilde{a}+\alpha \tilde{a}^2+\beta \tilde{a}^3$. The region is determined by two requirements: $i)$ the proper distance $d_{\rm P}(\tilde{z})$ agrees with the $\Lambda$CDM prediction (with Planck best-fit parameters) to within 3\% over the redshift range $0\leq \tilde{z}\leq 3$; $ii)$ the Hubble constant $\tilde{H}_0$ lies within the $2\sigma$ range from direct measurements. The blue shaded region is excluded by the $d_{\rm P}$ prior, the orange region is excluded by the $\tilde{H}_0$ prior, and the purple region represents the overlap. The red dots within the allowed region (labeled as Cases 1, 2 and 3) are three representative choices of coefficients for which we later derive observable predictions.}
\end{figure}

The next step is to determine the values of  $\alpha$ and $\beta$ for which our model provides a reasonable fit to data.
For starters we impose two conservative restrictions on the predicted expansion history. The first condition is that the proper distance $d_{\rm P}$,
normalized to $\tilde{H_0}$, agrees with the $\Lambda$CDM prediction to within 3\% over the redshift range $0\leq \tilde{z}\leq 3$.
Specifically, the quantity of interest is
\be
{\tilde H}_0 d_{\rm P}(\tilde z) = \int_0^{\tilde{z}}\rmd \tilde{z}'\frac{\tilde{H}_0}{\tilde{H}(\tilde{z}')}  \,.
\label{dphys}
\ee
The expression in $\Lambda$CDM cosmology is
\be
H_0^{\Lambda{\rm CDM}}d_{\rm P}^{\Lambda{\rm CDM}}(\tilde{z}) = \int_0^{\tilde{z}} \frac{{\rm d}\tilde{z}'}{\sqrt{\Omega_{\rm m}^{\Lambda{\rm CDM}} (1+\tilde{z}')^3 + 1- \Omega_{\rm m}^{\Lambda{\rm CDM}}}}\,,
\label{dphysLCDM}
\ee
where we will assume $\Omega_{\rm m}^{\Lambda{\rm CDM}} = 0.315$, corresponding to the Planck best-fit value~\cite{Ade:2015xua}. The 3\%~constraint is a conservative requirement that ensures good agreement with low-redshift geometric tests, such as Type Ia supernovae (Sec.~\ref{lumdis}) and Baryonic Acoustic Oscillations (Sec.~\ref{BAO}).

The second restriction pertains to the normalization of the cosmic ladder, set by the angular diameter distance to the CMB at $\tilde{z}_{\rm CMB} \simeq 1090$:
\be
d_{\rm A}(\tilde{z}_{\rm CMB}) = \frac{1}{1+\tilde{z}_{\rm CMB}}d_{\rm P}(\tilde{z}_{\rm CMB}) \,.
\label{dAus}
\ee
In $\Lambda$CDM cosmology with $\Omega_{\rm m}^{\Lambda{\rm CDM}} =0.315$, the result is $d_{\rm A}^{\Lambda {\rm CDM}}({\rm CMB}) \simeq \frac{3\times 10^{-3}}{H_0^{\Lambda{\rm CDM}}}$. Matching this to CMB data, the Planck mission can indirectly determine the Hubble constant, with result~\cite{Aghanim:2016yuo}:
$H_0^{\Lambda{\rm CDM}} = 66.93\pm 0.62~{\rm km}\;{\rm s}^{-1}{\rm Mpc}^{-1}$. This is well-known to be in tension (at the $\gsim\; 3 \sigma$ level) with the direct estimate with the Hubble Space Telescope~\cite{Riess:2016jrr} of $H_0^{\rm direct} = 73.24 \;\pm\; 1.74~{\rm km}\;{\rm s}^{-1}{\rm Mpc}^{-1}$. See~\cite{Bernal:2016gxb} for a nice discussion of this tension.

Similarly in our model we must match the predicted $d_{\rm A}(\tilde{z}_{\rm CMB})$ to Planck, which fixes $\tilde{H}_0$.\footnote{The CMB constraint on $H_0$ does not solely come from the sound horizon at last scattering, but also from the photon diffusion length scale which affects the Silk damping tale~\cite{Bashinsky:2003tk}. For the purpose of this preliminary analysis, we limit ourselves to matching $d_{\rm A}(\tilde{z}_{\rm CMB})$. We thank Adam Lidz for pointing this out to us.} Since our expansion history is generally different than $\Lambda$CDM, however, so is our value of $\tilde{H}_0$. We impose as a prior that our predicted $\tilde{H}_0$ lies within the range 65 to 75~${\rm km}\;{\rm s}^{-1}{\rm Mpc}^{-1}$. This range is chosen to include, at the lower end, the Planck best-fit $\Lambda$CDM value~\cite{Aghanim:2016yuo}, and, at the upper end, the direct Hubble Space Telescope (HST)~\cite{Riess:2016jrr} measurement. 

\begin{figure}[ht]
\centering
\begin{minipage}[b]{0.45\linewidth}
  \includegraphics[width=8 cm]{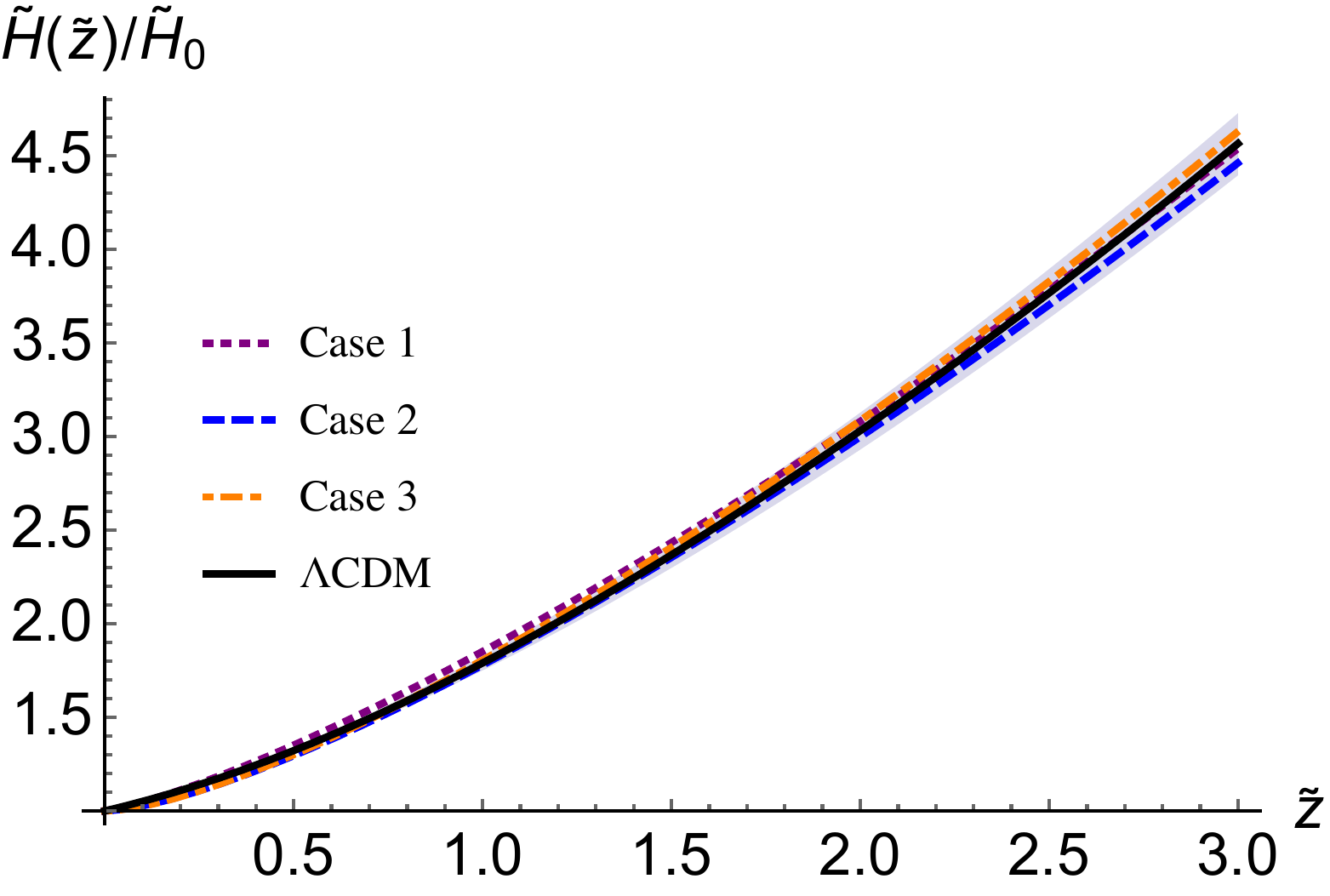} 
\end{minipage}
\qquad
\begin{minipage}[b]{0.45\linewidth}
  \includegraphics[width=8 cm]{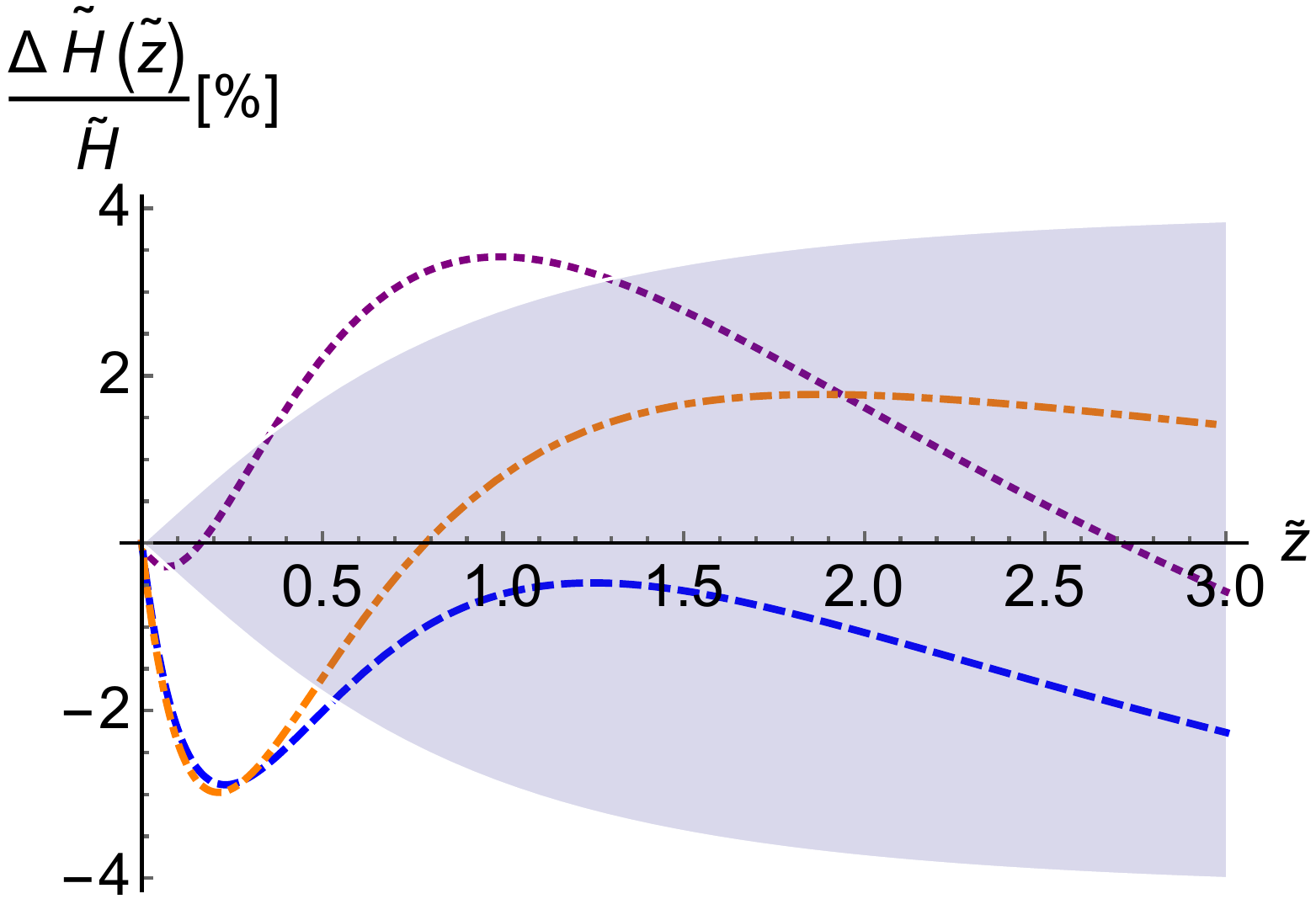}
\end{minipage}
\caption{\label{hubble}\textit{Left panel:} Normalized Hubble parameter as a function of the redshift, $\frac{\tilde{H}(\tilde{z})}{\tilde{H}_0}$, for our 3 fiducial sets of parameters. The grey band represents the Planck $\Lambda$CDM 1$\sigma$ range, $\Omega_{\rm m} = 0.315 \pm 0.026$, with the black curve corresponding to the central value. {\it Right Panel:} The fractional difference between our results and the $\Lambda$CDM prediction with $\Omega_{\rm m} = 0.315$, {\it i.e.}, $\frac{\Delta \tilde H}{\tilde H}\equiv \frac{\frac{\tilde{H}}{\tilde{H}_0} - \frac{H^{\Lambda{\rm CDM}}}{H_0^{\Lambda{\rm CDM}}}}{\frac{H^{\Lambda{\rm CDM}}}{H_0^{\Lambda{\rm CDM}}}}$.}
\end{figure}

The allowed region in $(\alpha,\beta)$ parameter space satisfying both requirements is shown as the white region in Fig.~\ref{alphabetaregion}. The blue shaded region is excluded by the $d_{\rm P}$ prior, the orange region is excluded by the $\tilde{H}_0$ prior, and the purple region represents the overlap. Within the white region we have selected 3 sample choices of coefficients, one central value (Case 2) and two extreme values (Cases 1 and 3):
\begin{align}
\nonumber
&\text{Case 1:} \qquad \qquad \alpha=-0.16\,, ~~~~\beta=-0.038\,;\\
&\text{Case 2:} \qquad \qquad \alpha=-0.08\,, ~~~~\beta=-0.09\,;\\
\nonumber
&\text{Case 3:} \qquad \qquad \alpha=-0.02\,, ~~~~\beta=-0.12\,.
\end{align}
These are shown as red dots within the allowed region. For illustrative purposes, in the remaining subsections we will calculate various observables and compare the results to the $\Lambda$CDM prediction for three sample choices of coefficients. For the record, the predicted Hubble constant in each case is
\begin{align}
\nonumber
&\tilde{H}_0^{\rm{Case\,1}}=74.1~{\rm km}\;{\rm s}^{-1}{\rm Mpc}^{-1}\,;\\
\nonumber
&\tilde{H}_0^{\rm{Case\,2}}=72.3~{\rm km}\;{\rm s}^{-1}{\rm Mpc}^{-1}\,; \\
&\tilde{H}_0^{\rm{Case\,3}}=67.5~{\rm km}\;{\rm s}^{-1}{\rm Mpc}^{-1}\,.
\label{Hubblevalues}
\end{align}

Figure~\ref{hubble} compares the normalized Hubble parameter as a function of the redshift, $\frac{\tilde{H}}{\tilde{H}_0}$, in each case together with the $\Lambda$CDM prediction $\frac{H^{\Lambda{\rm CDM}}}{H_0^{\Lambda{\rm CDM}}}$, over the redshift range $0\leq \tilde{z}\leq 3$. The grey band represents the Planck $\Lambda$CDM 1$\sigma$ range, $\Omega_{\rm m} = 0.315 \pm 0.026$, with the black curve corresponding to the central value. The right panel shows that in all cases the difference with $\Lambda$CDM is $\lsim\; 3$\% over this range.

\begin{figure}[ht]
\centering
\begin{minipage}[b]{0.45\linewidth}
  \includegraphics[width=8 cm]{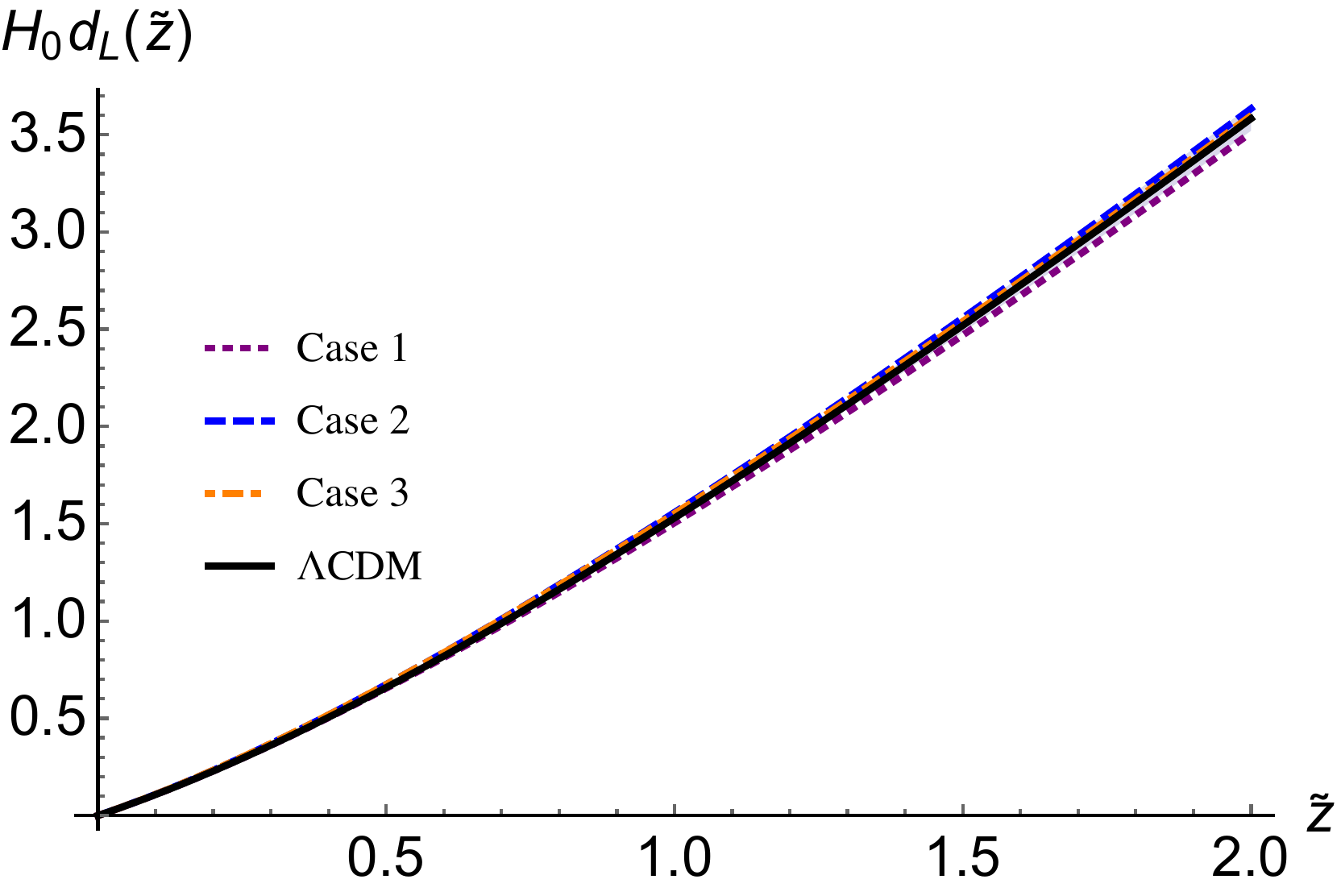} 
\end{minipage}
\qquad
\begin{minipage}[b]{0.45\linewidth}
  \includegraphics[width=8 cm]{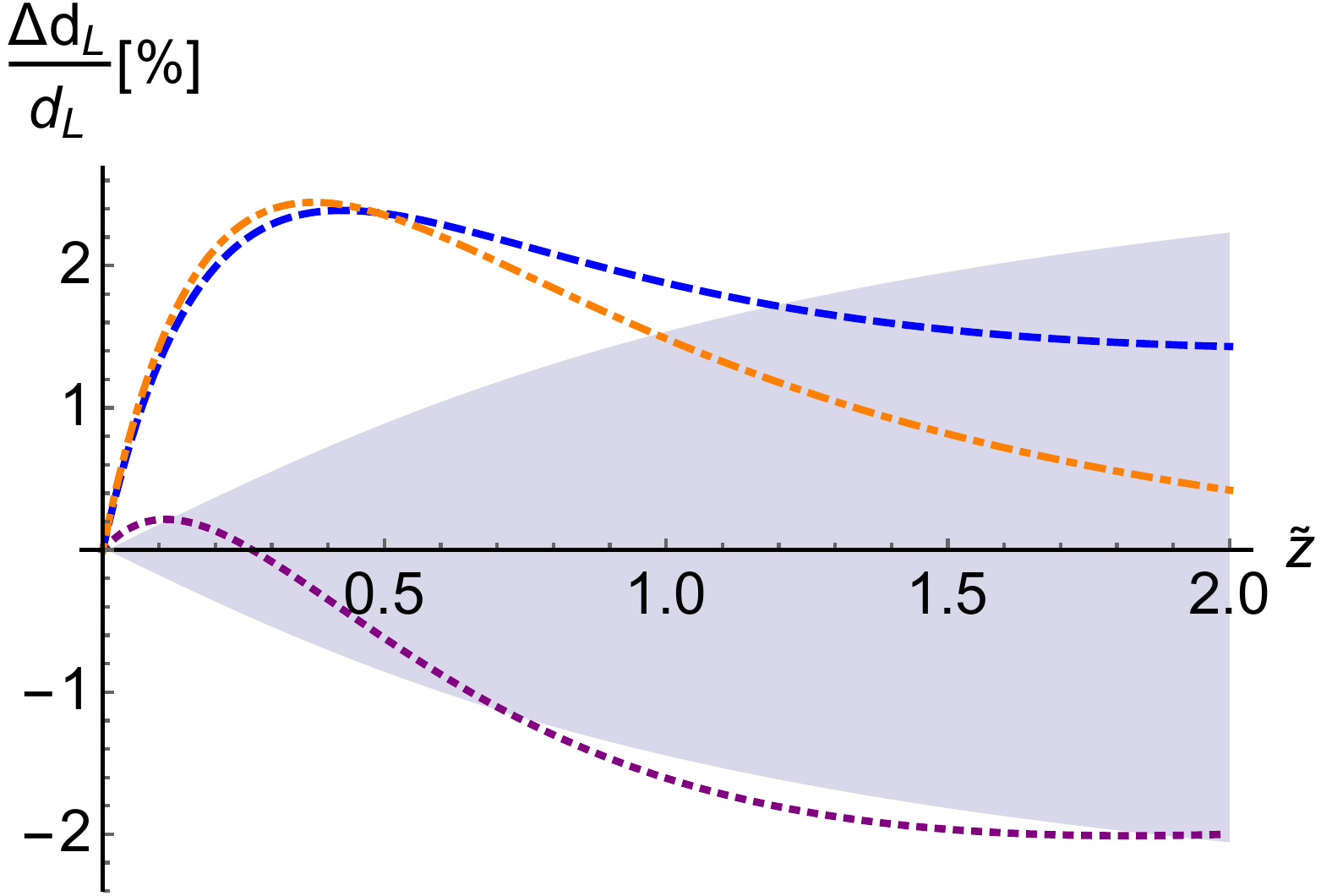}
\end{minipage}
\caption{\label{Lumdist}\textit{Left panel:} The luminosity distance as a function of redshift, with the same conventions as in Fig.~\ref{hubble}.{\it Right Panel:} The fractional difference with $\Lambda$CDM, defined as $\frac{\Delta d_{\rm L}}{d_{\rm L}}\equiv \frac{{\tilde H}_0 d_{\rm L}(\tilde z)-H_0^{\Lambda{\rm CDM}}d_{\rm L}^{\Lambda{\rm CDM}}(\tilde{z})}{H_0^{\Lambda{\rm CDM}}d_{\rm L}^{\Lambda{\rm CDM}}(\tilde{z})}$.}
\end{figure}

\subsection{Luminosity distance}
\label{lumdis}

Consider the luminosity distance $d_{\rm L}(\tilde z)$ to Type Ia supernovae (SNIa). This is related to the physical distance~\eqref{dphys} as usual by
\be
d_{\rm L}(\tilde z) = (1+\tilde z) d_{\rm P}(\tilde z) \,.
\ee
Type Ia observations offer tight constraints on $d_{\rm L}(\tilde z)$ over the redshift range $0\;\lsim\; \tilde{z}\;\lsim\; 1.5$~\cite{Betoule:2014frx}. An important constraint on our model is that our luminosity distance ${\tilde H}_0 d_{\rm L}$ agrees well with the corresponding $\Lambda$CDM expression $H_0^{\Lambda{\rm CDM}}d_{\rm L}^{\Lambda{\rm CDM}}$
over this redshift range. Figure~\ref{Lumdist} compares our model predictions for the three fiducial cases listed above and the $\Lambda$CDM prediction over the redshift range $0\leq \tilde{z}\leq 2$. The right panel shows that in all cases the difference with $\Lambda$CDM is $\lsim\; 2$\% over the entire range.

\begin{figure}[ht]
\centering
\begin{minipage}[b]{0.45\linewidth}
  \includegraphics[width=8 cm]{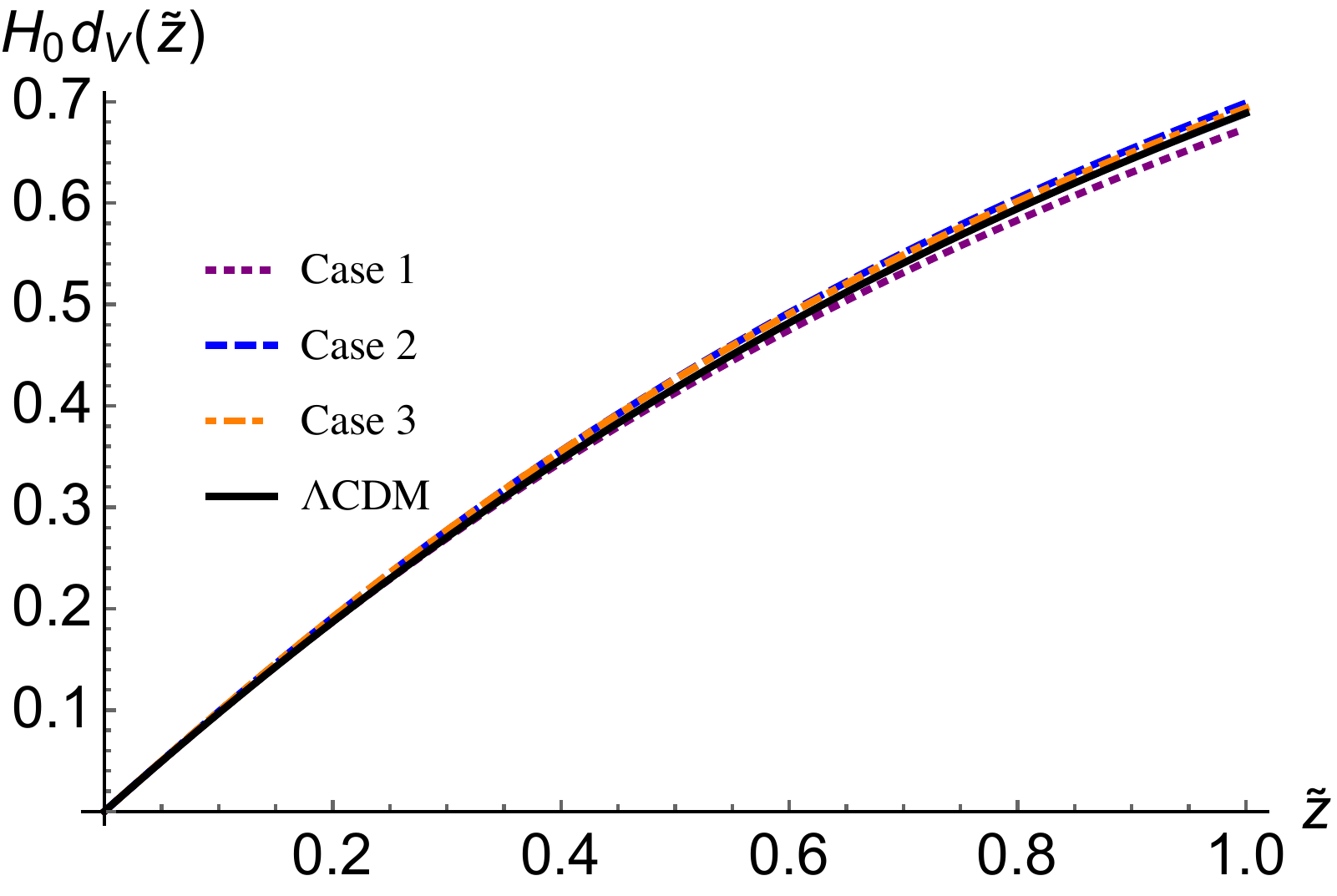} 
\end{minipage}
\qquad
\begin{minipage}[b]{0.45\linewidth}
  \includegraphics[width=8 cm]{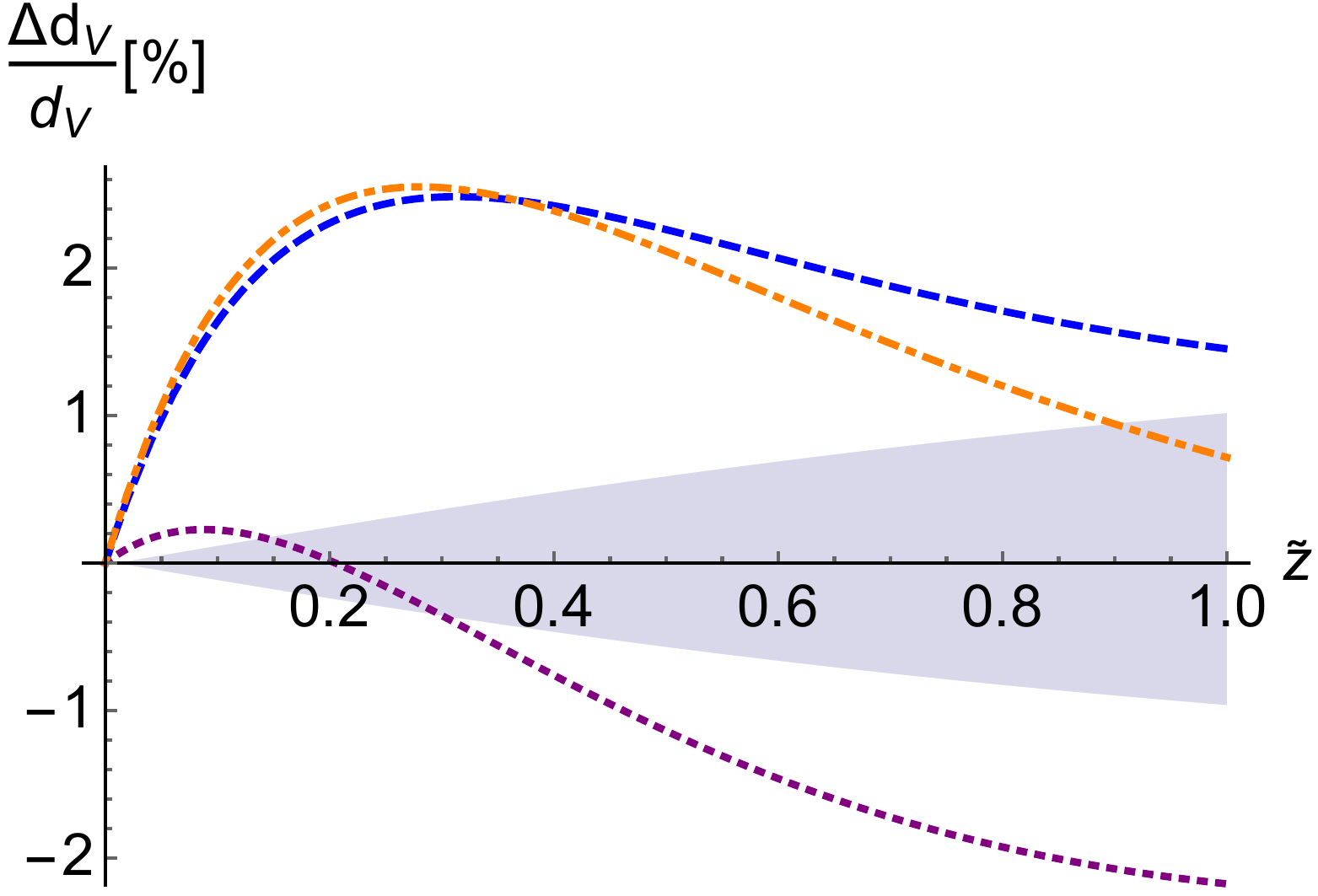}
\end{minipage}
\caption{\label{BAOdist}\textit{Left panel:} The distance relation $d_{\rm V}(\tilde{z})$ probed by BAO observations, with the same conventions as in Fig.~\ref{hubble}.{\it Right Panel:} The fractional difference with $\Lambda$CDM, defined as $\frac{\Delta d_{\rm V}}{d_{\rm V}}\equiv \frac{{\tilde H}_0 d_{\rm V}(\tilde z)-H_0^{\Lambda{\rm CDM}}d_{\rm V}^{\Lambda{\rm CDM}}(\tilde{z})}{H_0^{\Lambda{\rm CDM}}d_{\rm V}^{\Lambda{\rm CDM}}(\tilde{z})}$.}
\end{figure}

\subsection{Baryon Acoustic Oscillations}
\label{BAO}

Observations of Baryon Acoustic Oscillations (BAO) in large-scale structure surveys constrain the combination $r_{\rm drag}/d_{\rm V}(\tilde{z})$, where $r_{\rm drag}$ is the comoving sound horizon at the end of the baryon drag epoch~\cite{Eisenstein:1997ik}, and $d_{\rm V}$ is given by
\be
d_{\rm V}(\tilde{z}) = \left(d_{\rm P}^2(\tilde{z}) \frac{\tilde{z}}{\tilde{H}(\tilde{z})}\right)^{1/3}\,.
\ee
More precisely, galaxy surveys now have the statistics to decompose transverse and line-of-sight clustering information, thereby placing constraints on $d_{\rm A}$ and $H$ separately~\cite{Beutler:2016ixs}. However, for the purpose of this preliminary analysis we will contend ourselves with the comparison to the angle-averaged observable, $d_{\rm V}$. Various surveys constrain $r_{\rm drag}/d_{\rm V}(\tilde{z})$ to within 5-10\% percent of the $\Lambda$CDM best-fit prediction from Planck over the redshift range $0 \;\lsim\; \tilde{z}\;\lsim\; 1$, as summarized nicely in Fig.~14 of~\cite{Ade:2015xua}. Figure~\ref{BAOdist} compares our model predictions for $\tilde{H}_0d_{\rm V}(\tilde{z})$ for our 3 fiducial parameter sets with the $\Lambda$CDM prediction $H_0^{\Lambda{\rm CDM}}d_{\rm V}^{\Lambda{\rm CDM}}$ for $0\leq \tilde{z}\leq 1$. The right panel shows that in all cases the difference with $\Lambda$CDM is $\lsim\; 3$\% over this range.

\begin{figure}[ht]
\centering
\begin{minipage}[b]{0.45\linewidth}
  \includegraphics[width=8 cm]{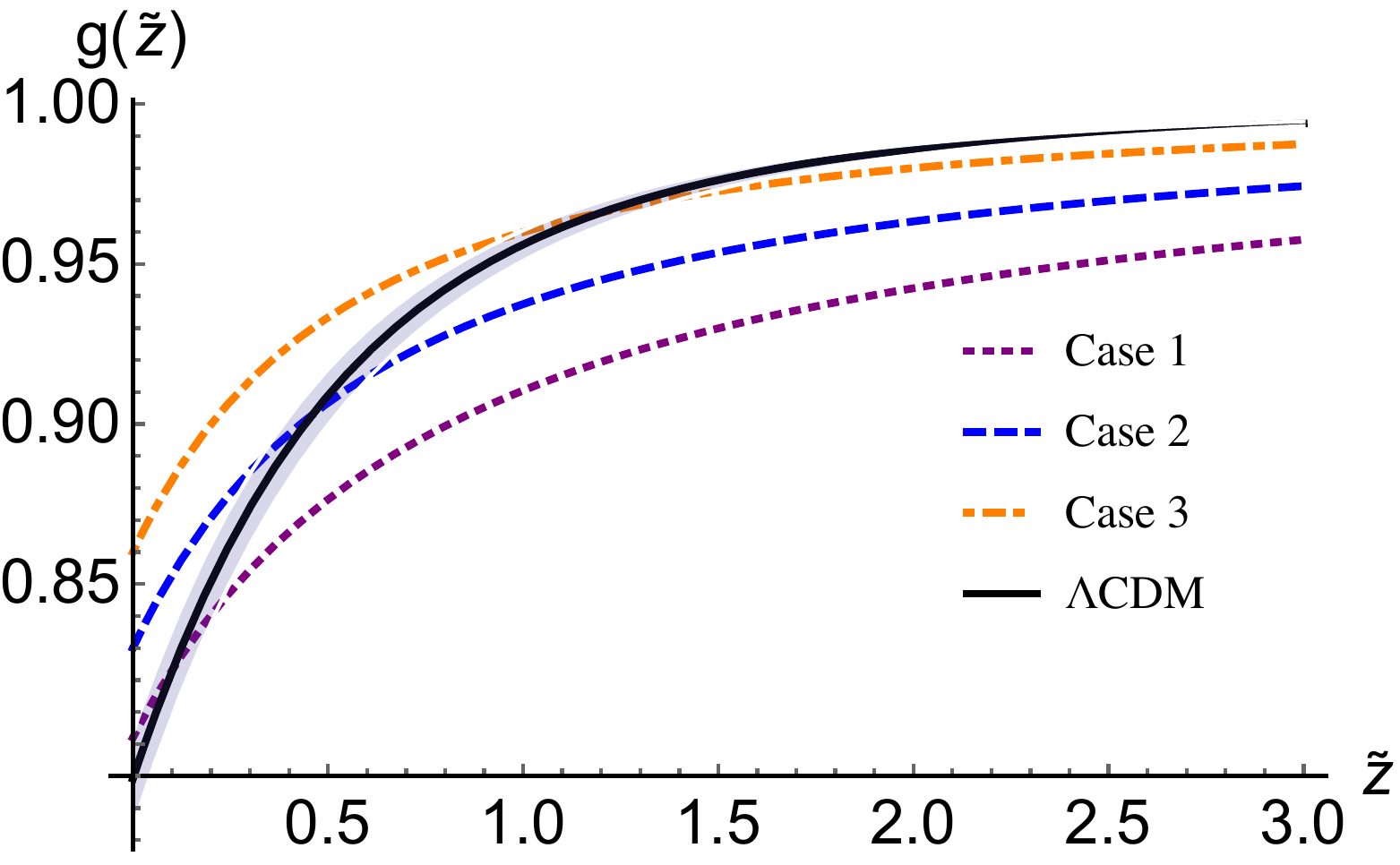} 
\end{minipage}
\qquad
\begin{minipage}[b]{0.45\linewidth}
  \includegraphics[width=8 cm]{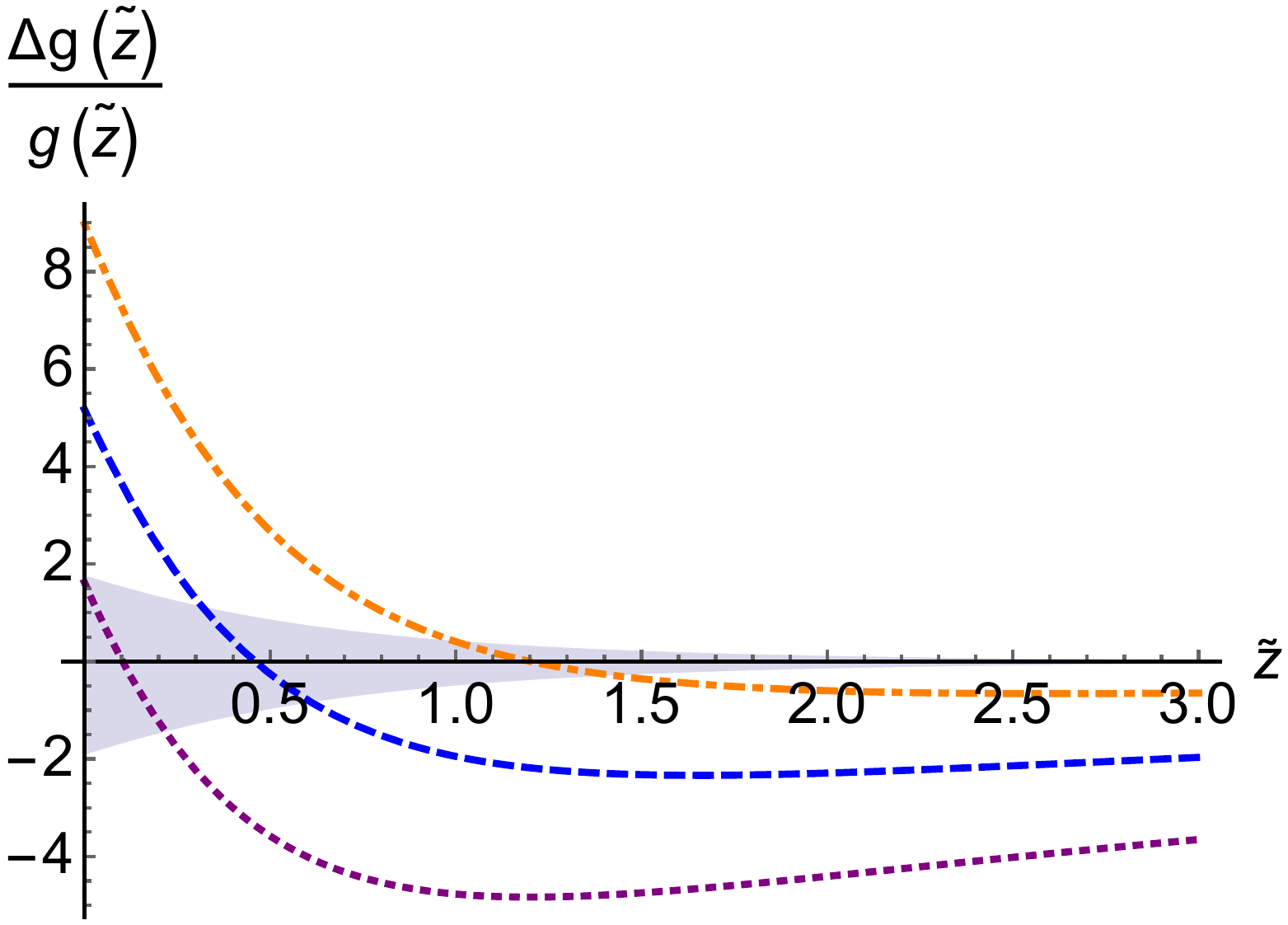}
\end{minipage}
\caption{\label{growthfactor}The rescaled DM growth factor, $g(\tilde{z})$, defined in~\eqref{growth}, is plotted as a function of redshift for the 3 fiducial sets of parameters, and compared to the 
$\Lambda$CDM prediction, calculated from Linder's fitting formula~\eqref{Linder}.}
\end{figure}

\subsection{Growth history of linear perturbations}
\label{growthhistorysection}

Next we consider the growth history of DM linear perturbations. For simplicity let us ignore the contribution from baryons, in which case
the function $g(\tilde{a})$ given by~\eqref{growth} matches the rescaled growth function defined in~\cite{Linder:2005in}.
Figure~\ref{growthfactor} plots this growth factor as a function of redshift for our three sets of fiducial parameters, together with the
$\Lambda$CDM fitting function proposed by~\cite{Linder:2005in}:
\be
g_{\Lambda{\rm CDM}}= e^{\int_0^a{\rm d}\ln a'~\left[ \Omega_{\rm m}(a')^{0.545}-1 \right]}\,.
\label{Linder}
\ee
The predicted $\sigma_8$ in each case is
\begin{align}
\nonumber
&\sigma_8^{\rm{Case\,1}}=0.84\,;\\
\nonumber
&\sigma_8^{\rm{Case\,2}}=0.87\,; \\
&\sigma_8^{\rm{Case\,3}}=0.90\,.
\end{align}
Clearly Case 1 offers the closest match to the $\Lambda$CDM best-fit normalization $\sigma_8=0.831 \pm 0.013 $~\cite{Ade:2015xua}. 
Interestingly,
recall from~\eqref{Hubblevalues} that this case also predicts the largest value of the Hubble constant, $\tilde{H}_0 =  74.1~{\rm km}\;{\rm s}^{-1}{\rm Mpc}^{-1}$,
in good agreement with direct estimates. Conversely, the Hubble constant for Case 3 is closest to the Planck $\Lambda$CDM value, but it is clear
from Fig.~\ref{growthfactor} that this case overpredicts the growth of structures.  

\begin{figure}[ht]
\centering
\includegraphics[width=8 cm]{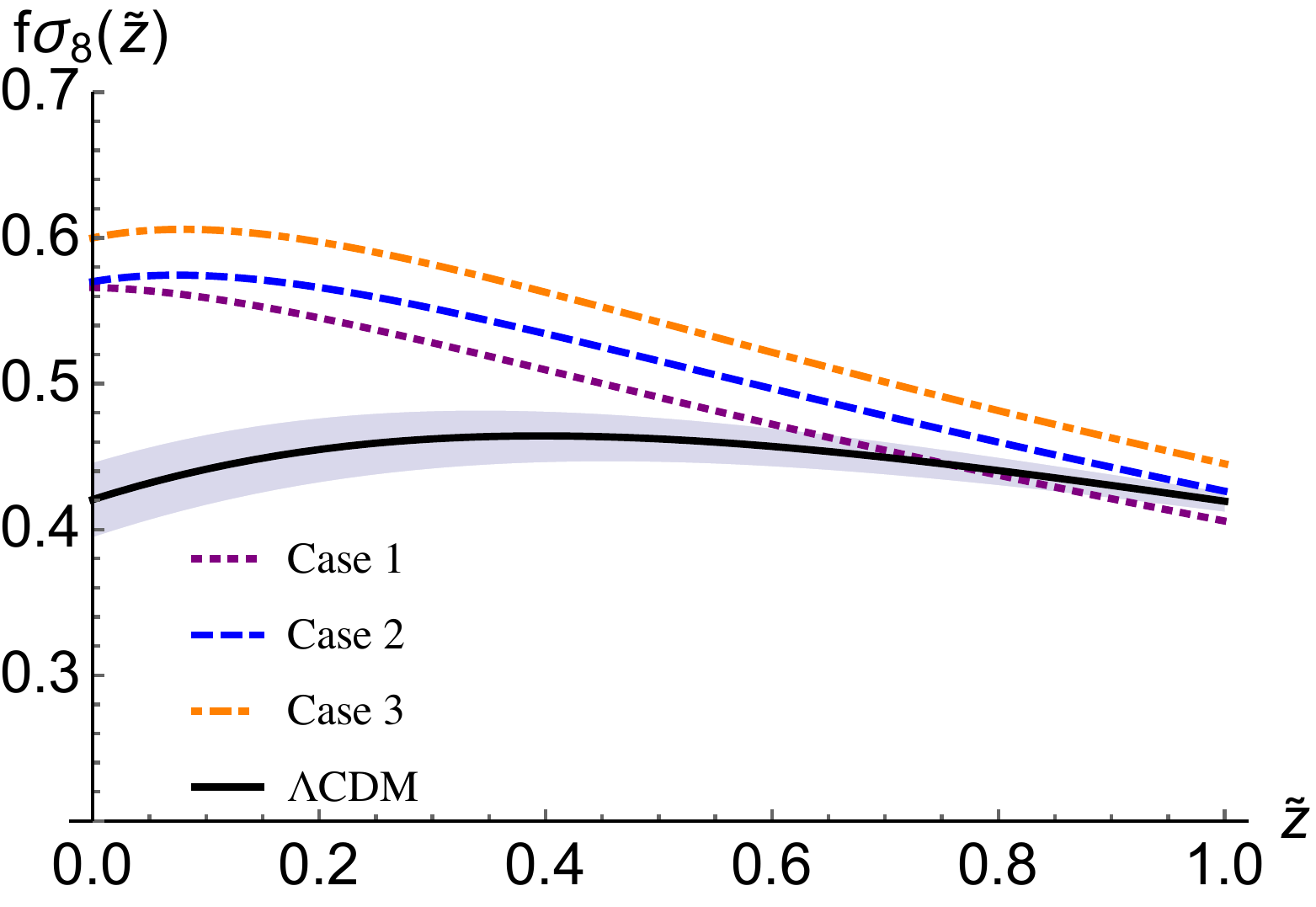} 
\caption{\label{fsigma8} The observable $f\sigma_8(\tilde{z})$, which is constrained by redshift-space distortions measurements, is plotted for our model and the $\Lambda$CDM.}
\end{figure}

Redshift-space distortions, {\it e.g.},~\cite{Beutler:2013yhm,Satpathy:2016tct}, constrain the combination $f\sigma_8$, where $f$ is the growth rate defined in~\eqref{fdef}. This quantity is plotted in Fig.~\ref{fsigma8}, together with the $\Lambda$CDM prediction. Our model agrees well with $\Lambda$CDM  at $\tilde{z}\sim1$, but we systematically predict a larger $f\sigma_8$ at low redshift. This seems in tension with the most recent constraints from the completed SDSS-III BOSS survey~\cite{Satpathy:2016tct}, which measured $f\sigma_8 = 0.430 \pm 0.054$ at $z_{\rm eff} = 0.38$, $f\sigma_8 =  0.452 \pm 0.057$ at $z_{\rm eff} = 0.51$, and $f\sigma_8 = 0.457 \pm 0.052$ at $z_{\rm eff} = 0.61$. We leave a more detailed comparison to redshift-space observations to future work.

 \subsection{Integrated Sachs-Wolfe effect}
\label{ISWsection}

One observable that is potentially problematic for us is the Integrated Sachs-Wolfe (ISW) effect, both through its impact on the low-multipole tail of the CMB power spectrum and 
on the cross-correlation with galaxy surveys. The ISW signal is determined by the rate of change of the effective gravitational potential $\Phi + \Psi$ felt by photons integrated along the null trajectory:
\be
\left(\frac{\delta T}{T}\right)_{\rm ISW} = \int_{\tau_{\rm rec}}^{\tau_0} {\rm d}\tau \frac{\partial}{\partial\tau} (\Phi + \Psi)\,,
\label{ISW}
\ee
where $\tau_{\rm rec}$ and $\tau_0$ are respectively the conformal time at recombination and at present. 

In our case the scalar potentials experienced by photons are those of the physical metric $\tilde{g}_{\mu\nu}$, {\it i.e.},
\be
{\rm d}\tilde{s}^2 = - (1 + 2\tilde{\Phi}){\rm d}t^2 + (1 - 2\tilde{\Psi})\tilde{a}^2{\rm d}\vec{x}^2\,.
\ee
In the maximally-disformal case of interest ($Q=1$), they are related to the Einstein-frame potentials by
\bea
\nonumber
\tilde{\Phi} &=& \Phi\,;\\
\tilde{\Psi} &=& \Psi - \frac{\delta R}{R} = \Psi - \frac{{\rm d}\ln R}{{\rm d}\ln X} \frac{\delta X}{X}\,.
\eea
where we have used the fact that the spatial metrics are related by $\tilde{g}_{ij} = R^2(X)g_{ij}$, and expanded $R(X)$ to linear order. 
An argument similar to that given for the conformal case in Sec.~\ref{confcouplingruledout} leads us to conclude that $\frac{{\rm d}\ln R}{{\rm d}\ln X} \sim \frac{1}{c_{\rm DM}^2}$ at late times in order for the $R(X)$ coupling to have a significant impact on the expansion history. Similarly it follows from the argument given around~\eqref{Xvary} that $ \frac{\delta X}{X} \sim c_{\rm DM}^2 \frac{\delta\rho_{\rm DM}}{\rho_{\rm DM}}$. Putting everything together, our ISW potential is
\be
\tilde{\Phi} + \tilde{\Psi} = \Phi + \Psi + {\cal O}(1) \frac{\delta\rho_{\rm DM}}{\rho_{\rm DM}}\,,
\label{ISWpot}
\ee
and thus receives a contribution proportional to $\frac{\delta\rho_{\rm DM}}{\rho_{\rm DM}}$. (It is worth emphasizing that~\eqref{ISWpot} only holds at late times, when $R$ is significantly different from unity and drives cosmic acceleration.) 

The issue comes from the fact that $\frac{\delta\rho_{\rm DM}}{\rho_{\rm DM}} \sim k^2 \Phi$ is strongly scale-dependent and peaks on small scales. This naively implies a large ISW signal, which may be problematic. On the other hand, it is worth noting that observations favor (at the $\simeq 2\sigma$ level) a larger ISW effect than predicted by standard $\Lambda$CDM cosmology~\cite{Liu:2015xfa}. Meanwhile, the contribution from $\Phi + \Psi$ is expected to be small. Indeed, at least to the extent that baryons are negligible, the (Einstein-frame) gravitational potentials experience a matter-dominated universe to the present time, and hence are constant on linear scales. Therefore only the $\delta\rho_{\rm DM}/\rho_{\rm DM}$ term is expected to contribute significantly to~\eqref{ISW}. A quantitative treatment of the ISW effect, together with detailed predictions for the CMB and matter power spectrum, is currently in progress and will be presented in a future paper~\cite{AdamVin}.

Another observable that is determined by the combination $\Phi + \Psi$ is the large-scale weak lensing power spectrum measured by Planck.\footnote{We thank Paolo Creminelli for discussions of this point.} This could similarly be affected by the density-dependent contribution, though its impact will ultimately depend on the behavior of $R(X)$ where the lensing kernel peaks. This also deserves further study.

\section{Discussion}

In this paper we presented a third avenue for generating cosmic acceleration, without a source of negative pressure and without new degrees of freedom beyond those of Einstein gravity. The mechanism relies on the coupling between DM and baryons through an effective metric. Dark matter couples to an Einstein-frame metric, and experiences a matter-dominated, decelerating cosmology up to the present time. Ordinary matter, meanwhile, couples to an effective metric that depends both on the Einstein-frame metric and on the DM density. By construction this effective metric reduces to the Einstein-frame metric at early times, but describes an accelerating cosmology at late times.

Linear perturbations are stable and propagate with arbitrarily small sound speed, at least in the case of maximally-disformal or pressure coupling $Q = 1$. The case of conformal coupling, on the other hand, generically results in a relativistic sound speed at late times, and is therefore observationally disfavored~\cite{Sandvik:2002jz}. As the name suggests, the case of pressure coupling $Q = 1$ implies that pressureless sources ({\it i.e.}, non-relativistic particles) are in fact decoupled from the DM. The DM only affects relativistic particles, in particular all observational consequences derive from the effect of the ambient DM background on the propagation of photons. In this sense our proposal is spiritually similar to the old idea of ``tired light", proposed long ago by Zwicky as an alternative to the expanding universe. In our case the {\it accelerating} universe is a consequence of photons interacting with the DM medium along the line of sight. 

We do not claim that our model is somehow better-motivated from a particle physics standpoint than existing explanations for cosmic acceleration. 
After all we have at our disposal {\it a prior} two free functions $Q(X)$ and $R(X)$ (one function in the maximally-disformal case) that must be engineered to reproduce standard evolution at early times and generate an accelerating universe at late times. This is similar to the tuning inherent to quintessence models, where one specifies a scalar potential
to obtain the desired evolution. Nevertheless the mechanism is sufficiently novel and different than existing explanations on the market that it is definitely worth exploring its
observational consequences.

As a sanity check, we have performed a preliminary check for a few key cosmological observables, focusing on the maximally-disformal coupling, and
compared the results to $\Lambda$CDM predictions. For a simple parametrization of the $R(X)$ coupling function, our model can successfully reproduce various geometric constraints, including the luminosity distance relation and BAO measurements. For density perturbations, our model predicts an intriguing connection between the growth factor and the Hubble constant (which is fixed by matching the angular diameter distance to the CMB). To get a growth history similar to the $\Lambda$CDM prediction, our model predicts a higher
$H_0$, closer to the value preferred by direct estimates. On the flip side, we tend to overpredict the growth of structures whenever $H_0$ is comparable
to the Planck preferred value. 

One observable that may be problematic is the ISW effect, both through its impact on the CMB power spectrum at low multipoles and 
on the cross-correlation with galaxy surveys. The form of our coupling implies a density-dependent contribution to this observable,
which may yield too large a signal on small scales. On the other hand, as mentioned already, there is a $2\sigma$ excess in the observed
cross-correlation relative to the $\Lambda$CDM prediction~\cite{Liu:2015xfa}. In ongoing work we are modifying the CAMB code to
calculate the CMB and matter power spectra. This will allow us to make rigorous predictions and check, in particular, whether the ISW signal
is compatible with observations.

\noindent {\bf Acknowledgements:} We thank Paolo Creminelli, Gia Dvali, Brian Henning, Wayne Hu, Bhuvnesh Jain, Austin Joyce, David E.~Kaplan, Jared Kaplan, Daliang Li, Adam Lidz, Vinicius Miranda, David Poland, Adam Solomon and Mark Trodden for helpful discussions. L.B. is supported by US Department of Energy grant \text{DE-SC0007968}. J.K. is supported in part by NSF CAREER Award PHY-1145525, NASA ATP grant NNX11AI95G, and the Charles E. Kaufman Foundation of the Pittsburgh Foundation. The work of J.W. is supported by Simons Foundation award $\#$488651.

\section*{Appendix: Equivalence Between Fluid Descriptions}
\renewcommand{\theequation}{A-\Roman{equation}}
\setcounter{equation}{0} 

In this Appendix we elaborate on the points briefly mentioned in Sec.~\ref{setupsection} regarding the equivalence of DM effective theories in the hydrodynamical regime.
In particular we clarify the physical implications of neglecting DM vorticity.

The most general effective field theory description of a fluid/solid continuum includes not only the longitudinal mode but also the transverse degrees of freedom. Specifically, following \cite{Dubovsky:2005xd, Wang:2013ic} a fluid/solid is described by 3 Lorentz scalars $\phi^I(x^\mu)$, $I = 1,2,3$, specifying the comoving position of each fluid element as a function of laboratory space-time coordinates $x^\mu$. 
For a homogeneous and isotropic fluid/solid, the action should be invariant under internal translations $\phi^I\rightarrow \phi^I + a^I$ and rotations $\phi^I\rightarrow R^I_{~J}\phi^J$.
Furthermore, in the case of a perfect fluid, shear deformations come at no cost in energy, hence the action should also be invariant under volume-preserving diffeomorphisms:
\be
\phi^I \rightarrow \hat{\phi}^I\;;\qquad \det \frac{\partial \hat{\phi}^I}{\partial \phi^J} = 1\,.
\ee
At lowest number in derivatives, this implies that the action is a function of the determinant:
\be
{\cal L}_B = -\sqrt{-g} \rho(B)\,;\qquad B \equiv \det \left(g^{\mu\nu}\partial_\mu\phi^I \partial_\nu\phi^J\right)\,.
\label{B Action}
\ee
The $\phi^I$'s have units of length, hence $B$ is dimensionless. 

The equation of motion following from this action reads
\be
\pd_\mu\bigg(\sqrt{-g}\rho_{,B}(B) B (B^{-1})_{IJ}g^{\mu\nu}\pd_\nu\phi^J\bigg)=0\;.
\ee
Therefore, an isolated fluid (no external source) in the flat spacetime ($g_{\mu\nu}=\eta_{\mu\nu}$) allows the following ground state configuration: 
\be
\bar{\phi}^I = \alpha x^I\;,
\ee
with $\alpha$ being a dimensionless constant. We parameterize the fluctuations around this ground state by
\be
\pi^I=\phi^I -\bar{\phi}^I\;.
\ee
It is straightforward to show that the stress tensor of this action can be cast into a form of a perfect fluid
\be\label{T Fluid}
T_{\mu\nu}=(\rho+P)u_\mu u_\nu + P g_{\mu\nu}\,,
\ee
where the energy density $\rho$ and pressure $P$ are given by
\be
\rho(\phi)= \rho(B)\;,\quad P(\phi)=2 B \rho_{,B}(B)-\rho(B)\,,
\ee
and the velocity field $u^{\mu}$ by
\be
u^\mu(\phi)=\frac{1}{6\sqrt{B}}\epsilon^{\mu\alpha\beta\gamma}
\epsilon_{IJK} \pd_\alpha\phi^I\pd_\beta\phi^J\pd_\gamma\phi^K\,.
\ee
The three degrees of freedom can be regrouped into a longitudinal phonon field and a vortex field. At {\it linear} order in fluctuations, they are just the longitudinal and transverse parts of the vector $\pi^I$.

When there is no vertex field excited, the number of degrees of freedom reduces to one. Therefore it is not surprising that the fluid action~\eqref{B Action} enjoys a simpler dual description involving only one scalar field:
\be
{\cal L}_X=\sqrt{-g}P(X)\;,\quad X=-\pd_\mu\Theta \pd_\nu \Theta g^{\mu\nu}\,.
\ee
We briefly review the proof of the equivalence between two effective descriptions of fluids. The stress tensor of the action ${\cal L}_X$ also takes the perfect fluid form~\eqref{T Fluid}, with
\begin{align}\label{Duality}
\rho(\Theta)=2P_{,X}(X)X-P(X)\;,\quad P(\Theta)= P(X)\;,\quad u_\mu(\Theta)=-\frac{1}{\sqrt{X}}\pd_\mu\Theta\,.
\end{align}
\textit{The precise statement of the equivalence between $\rho(B)$ and $P(X)$ description says that, one can establish some relation between $\phi^I$ and $\Theta$, such that
\be
\rho(\phi)=\rho(\Theta)\;;\quad P(\phi)=P(\Theta)\;;\quad u_{\mu}(\phi)=u_\mu(\Theta)\,.
\ee
}
We will prove this by constructing an explicit map. 

The key point of this construction lies in finding the correct condition to eliminate the extra degree of freedom present in the $\rho(B)$ language. It turns out that this is possible in the absence of vorticity. Let us begin by defining the following form
\be
{\bf V}=V_\mu {\rm d} x^\mu = -\rho_{,B}(B)\sqrt{B} u_\mu(\phi) {\rm d} x^\mu\,,
\label{V one form}
\ee
then the relativistic version of vorticity of the fluid is given by the two form $d {\bf V}$. Vanishing vorticity implies that the one-from $\bf V$ must be closed, thus
\be\label{No Vorticity}
V_\mu= -\rho_{,B}(B)\sqrt{B} u_\mu(\phi)=\Lambda^4 \pd_\mu\Theta\;.
\ee
The proportionality constant is introduced to match dimensions on both sides.

It follows immediately from~\eqref{No Vorticity} that 
\be
X=B f_{,B}(B)^2\;, \quad f(B)\equiv -\Lambda^{-4}\rho(B)\,.
\ee
Moreover, if one define $P(X)=\Lambda^4 p(X)$ with 
\be\label{P in F}
p\big(X(B)\big)=f(B)-2B f_{,B}(B)\;,
\ee
then all the equalities in~\eqref{Duality} are satisfied. That completes our construction.

To get some intuition about this duality, let us work out the relation between fluctuation $\phi$ and $\Theta$. Writing $\Theta=c t+\theta$,~\eqref{No Vorticity} becomes
\be
(c+\dot{\theta}, \pd_i \theta)=-f^0_{,B}
\left[
1+\left(1+\frac{2f^0_{,BB}}{f^0_{,B}}\right)\pd_I\pi^I
\right]
(1,\pi^I)+{\cal O}(\pi^2)\;,
\ee
where we have rescaled the coordinate in such a way that $\alpha=1$ and $B^0=\alpha^3=1$, and denoted by $f^{(n)}_0=f^{(n)}(1)\;.$ Therefore one obtains that
\begin{align}
c&=-f^0_{,B} \;,\\
c \dot{\pi^I}&=\pd_I \theta\;,\\
\dot{\theta}&=c\left(1+\frac{2 f^0_{,BB}}{f^0_{,B}}\right)\pd_I \pi^I\;.
\end{align}
The second equation implies that $\pi^I$ must be a gradient mode $\pi^I=\frac{\pd^I}{\sqrt{-\pd^2}}\pi_L$\;, while the second and third equation combined require that 
\be
\ddot{\pi}_L=\left(1+\frac{2 f^0_{,BB}}{f^0_{,B}}\right)\pd^2 \pi_L=c_s^2 \pd^2 \pi_L\;,
\ee
which is nothing but the linearized equation of motion for longitudinal $\pi^I$. That is, the no-vorticity condition can only be satisfied for on-shell configurations; the duality between $\rho(B)$ and $P(X)$ is a classical equivalence. 

On the other hand, the above construction does not work for fluids coupled to external source. For instance, the action in $B$ language
\be
{\cal L}=\sqrt{-g}\bigg(-\rho(B)+\Omega(B) {\cal L}_{\rm baryon}\bigg)
\ee
is {\it not} simply equivalent to that in the $X$ language
\be\label{L prime}
{\cal L}'=\sqrt{-g}\bigg(P(X)+Q(X) {\cal L}_{\rm baryon}\bigg)\;,
\ee
where $P(X)$ and $\rho(B)$ are related by \eqref{P in F}. This is because in the presence of external source ({\it i.e.}, baryons in the above example), the vorticity is no longer conserved: $d {\bf V}\ne 0$\;. In order to establish the equivalence, one would need to find a new, conserved vortex vector $\hat{\bf V}$ by including baryon fields. For concreteness, we have chosen to formulate our theory in the $X$ language.

\end{document}